\documentclass[reprint,amsmath,amssymb,aps,prd,nofootinbib,superscriptaddress]{revtex4-2}

\usepackage{graphicx}
\usepackage{hyperref}

\providecommand{\apj}{ApJ}
\providecommand{\apjl}{ApJL}
\providecommand{\apjs}{ApJS}
\providecommand{\aap}{A\&A}

\providecommand{\mnras}{MNRAS}

\providecommand{\prl}{Phys.~Rev.~Lett.}
\providecommand{\prd}{Phys.~Rev.~D}
\providecommand{\physletb}{Phys.~Lett.~B}

\usepackage[dvipsnames]{xcolor}

\graphicspath{{figures/}}

\begin{document}

\title{Search for Axion-like Particles using Blazar Observations with the HAWC Observatory}

\author{R.~Alfaro}
\affiliation{Instituto de F\'{i}sica, Universidad Nacional Aut\'{o}noma de M\'{e}xico, Ciudad de Mexico, Mexico }
\author{C.~Alvarez}
\affiliation{Universidad Aut\'{o}noma de Chiapas, Tuxtla Guti\'{e}rrez, Chiapas, M\'{e}xico}
\author{A.~Andr\'{e}s}
\author{E.~Anita-Rangel}
\affiliation{Instituto de Astronom\'{i}a, Universidad Nacional Aut\'{o}noma de M\'{e}xico, Ciudad de Mexico, Mexico }
\author{M.~Araya}
\affiliation{Universidad de Costa Rica, San Jos\'{e} 2060, Costa Rica}
\author{J.C.~Arteaga-Vel\'{a}zquez}
\affiliation{Universidad Michoacana de San Nicol\'{a}s de Hidalgo, Morelia, Mexico }
\author{D.~Avila Rojas}
\affiliation{Instituto de Astronom\'{i}a, Universidad Nacional Aut\'{o}noma de M\'{e}xico, Ciudad de Mexico, Mexico }
\author{H.A.~Ayala Solares}
\affiliation{Temple University, Department of Physics, 1925 N. 12th Street, Philadelphia, PA 19122, USA}
\author{R.~Babu}
\affiliation{Department of Physics and Astronomy, Michigan State University, East Lansing, MI, USA }
\author{A.~Bernal}
\affiliation{Instituto de Astronom\'{i}a, Universidad Nacional Aut\'{o}noma de M\'{e}xico, Ciudad de Mexico, Mexico }
\author{A.~Carrami\~{n}ana}
\affiliation{Instituto Nacional de Astrof\'{i}sica, \'{O}ptica y Electr\'{o}nica, Puebla, Mexico }
\author{F.~Carre\'{o}n}
\affiliation{Instituto de Astronom\'{i}a, Universidad Nacional Aut\'{o}noma de M\'{e}xico, Ciudad de Mexico, Mexico }
\author{S.~Casanova}
\affiliation{Institute of Nuclear Physics Polish Academy of Sciences, PL-31342 IFJ-PAN, Krakow, Poland }
\author{A.L.~Colmenero-Cesar}
\affiliation{Universidad Michoacana de San Nicol\'{a}s de Hidalgo, Morelia, Mexico }
\author{J.~Cotzomi}
\affiliation{Facultad de Ciencias F\'{i}sico Matem\'{a}ticas, Benem\'{e}rita Universidad Aut\'{o}noma de Puebla, Puebla, Mexico }
\author{N.~Di Lalla}
\affiliation{Department of Physics, Stanford University: Stanford, CA 94305--4060, USA}
\author{R.~Diaz Hernandez}
\affiliation{Instituto Nacional de Astrof\'{i}sica, \'{O}ptica y Electr\'{o}nica, Puebla, Mexico }
\author{M.A.~DuVernois}
\affiliation{Dept. of Physics and Wisconsin IceCube Particle Astrophysics Center, University of Wisconsin{\textemdash}Madison, Madison, WI, USA}
\author{T.~Ergin}
\affiliation{Department of Physics and Astronomy, Michigan State University, East Lansing, MI, USA }
\author{C.~Espinoza}
\affiliation{Instituto de F\'{i}sica, Universidad Nacional Aut\'{o}noma de M\'{e}xico, Ciudad de Mexico, Mexico }
\author{N.~Fraija}
\author{S.~Fraija}
\affiliation{Instituto de Astronom\'{i}a, Universidad Nacional Aut\'{o}noma de M\'{e}xico, Ciudad de Mexico, Mexico }
\author{J.A.~Garc\'{\i}a-Gonz\'{a}lez}
\affiliation{Tecnologico de Monterrey, Escuela de Ingenier\'{i}a y Ciencias, Ave. Eugenio Garza Sada 2501, Monterrey, N.L., Mexico, 64849}
\author{F.~Garfias}
\affiliation{Instituto de Astronom\'{i}a, Universidad Nacional Aut\'{o}noma de M\'{e}xico, Ciudad de Mexico, Mexico }
\author{N.~Ghosh}
\affiliation{Department of Physics, Michigan Technological University, Houghton, MI, USA }
\author{S.~Giri}
\affiliation{Institute of Nuclear Physics Polish Academy of Sciences, PL-31342 IFJ-PAN, Krakow, Poland }
\author{M.M.~Gonz\'{a}lez}
\affiliation{Instituto de Astronom\'{i}a, Universidad Nacional Aut\'{o}noma de M\'{e}xico, Ciudad de Mexico, Mexico }
\author{J.A.~Gonz\'{a}lez}
\affiliation{Universidad Michoacana de San Nicol\'{a}s de Hidalgo, Morelia, Mexico }
\author{J.A.~Goodman}
\affiliation{Department of Physics, University of Maryland, College Park, MD, USA }
\author{J.~Gyeong}
\affiliation{Department of Physics, Sungkyunkwan University, Suwon 16419, Republic of Korea}
\author{J.P.~Harding}
\affiliation{Los Alamos National Laboratory, Los Alamos, NM, USA }
\author{I.~Herzog}
\affiliation{Department of Physics and Astronomy, Michigan State University, East Lansing, MI, USA }
\author{J.~Hinton}
\affiliation{Max-Planck Institute for Nuclear Physics, 69117 Heidelberg, Germany}
\author{D.~Huang}
\affiliation{Department of Physics and Astronomy, University of Delaware, Newark, DE, USA}
\author{F.~Hueyotl-Zahuantitla}
\affiliation{Universidad Aut\'{o}noma de Chiapas, Tuxtla Guti\'{e}rrez, Chiapas, M\'{e}xico}
\author{S.~Kaufmann}
\affiliation{Universidad Politecnica de Pachuca, Pachuca, Hgo, Mexico }
\author{K.~Leavitt}
\affiliation{Department of Physics, Michigan Technological University, Houghton, MI, USA }
\author{J.S.~Lee}
\email[Contact author: ]{Jason.Lee@cern.ch}
\affiliation{Department of Physics, Yonsei University, Seoul, Republic of Korea}
\author{H.~Le\'{o}n Vargas}
\affiliation{Instituto de F\'{i}sica, Universidad Nacional Aut\'{o}noma de M\'{e}xico, Ciudad de Mexico, Mexico }
\author{J.T.~Linnemann}
\affiliation{Department of Physics and Astronomy, Michigan State University, East Lansing, MI, USA }
\author{A.L.~Longinotti}
\affiliation{Instituto de Astronom\'{i}a, Universidad Nacional Aut\'{o}noma de M\'{e}xico, Ciudad de Mexico, Mexico }
\author{G.~Luis-Raya}
\affiliation{Universidad Politecnica de Pachuca, Pachuca, Hgo, Mexico }
\author{O.~Martinez}
\affiliation{Facultad de Ciencias F\'{i}sico Matem\'{a}ticas, Benem\'{e}rita Universidad Aut\'{o}noma de Puebla, Puebla, Mexico }
\author{J.~Mart\'{\i}nez-Castro}
\affiliation{Centro de Investigaci\'on en Computaci\'on, Instituto Polit\'ecnico Nacional, M\'exico City, M\'exico.}
\author{H.~Mart\'{\i}nez-Huerta}
\affiliation{Departamento de F\'{\i}sica y Matem\'{a}ticas, Universidad de Monterrey, Av. Morones Prieto 4500, San Pedro Garza Garc\'{\i}a 66238, Nuevo Le\'{o}n, M\'{e}xico}
\author{J.A.~Matthews}
\affiliation{Dept of Physics and Astronomy, University of New Mexico, Albuquerque, NM, USA }
\author{P.~Miranda-Romagnoli}
\affiliation{Universidad Aut\'{o}noma del Estado de Hidalgo, Pachuca, Mexico }
\author{P.E.~Mir\'{o}n-Enriquez}
\affiliation{Instituto de Astronom\'{i}a, Universidad Nacional Aut\'{o}noma de M\'{e}xico, Ciudad de Mexico, Mexico }
\author{J.A.~Morales-Soto}
\affiliation{Universidad Michoacana de San Nicol\'{a}s de Hidalgo, Morelia, Mexico }
\author{E.~Moreno}
\affiliation{Facultad de Ciencias F\'{i}sico Matem\'{a}ticas, Benem\'{e}rita Universidad Aut\'{o}noma de Puebla, Puebla, Mexico }
\author{M.~Mostaf\'{a}}
\affiliation{Temple University, Department of Physics, 1925 N. 12th Street, Philadelphia, PA 19122, USA}
\author{M.~Najafi}
\affiliation{Department of Physics, Michigan Technological University, Houghton, MI, USA }
\author{L.~Nellen}
\affiliation{Instituto de Ciencias Nucleares, Universidad Nacional Aut\'{o}noma de Mexico, Ciudad de Mexico, Mexico }
\author{N.~Omodei}
\affiliation{Department of Physics, Stanford University: Stanford, CA 94305--4060, USA}
\author{E.~Ponce}
\affiliation{Facultad de Ciencias F\'{i}sico Matem\'{a}ticas, Benem\'{e}rita Universidad Aut\'{o}noma de Puebla, Puebla, Mexico }
\author{Y.~P\'{e}rez Araujo}
\affiliation{Instituto de F\'{i}sica, Universidad Nacional Aut\'{o}noma de M\'{e}xico, Ciudad de Mexico, Mexico }
\author{E.G.~P\'{e}rez-P\'{e}rez}
\affiliation{Universidad Politecnica de Pachuca, Pachuca, Hgo, Mexico }
\author{C.D.~Rho}
\affiliation{Department of Physics, Sungkyunkwan University, Suwon 16419, Republic of Korea}
\author{D.~Rosa-Gonz\'{a}lez}
\affiliation{Instituto Nacional de Astrof\'{i}sica, \'{O}ptica y Electr\'{o}nica, Puebla, Mexico }
\author{M.~Roth}
\affiliation{Los Alamos National Laboratory, Los Alamos, NM, USA }
\author{A.~Sandoval}
\author{J.~Serna-Franco}
\affiliation{Instituto de F\'{i}sica, Universidad Nacional Aut\'{o}noma de M\'{e}xico, Ciudad de Mexico, Mexico }
\author{M.~Shin}
\author{Y.~Son}
\email[Contact author: ]{youngwan.son@skku.edu}
\affiliation{Department of Physics, Sungkyunkwan University, Suwon 16419, Republic of Korea}
\author{R.W.~Springer}
\affiliation{Department of Physics and Astronomy, University of Utah, Salt Lake City, UT, USA }
\author{O.~Tibolla}
\affiliation{Universidad Politecnica de Pachuca, Pachuca, Hgo, Mexico }
\author{K.~Tollefson}
\affiliation{Department of Physics and Astronomy, Michigan State University, East Lansing, MI, USA }
\author{I.~Torres}
\author{F.~Ure\~{n}a-Mena}
\affiliation{Instituto Nacional de Astrof\'{i}sica, \'{O}ptica y Electr\'{o}nica, Puebla, Mexico }
\author{E.~Varela}
\affiliation{Facultad de Ciencias F\'{i}sico Matem\'{a}ticas, Benem\'{e}rita Universidad Aut\'{o}noma de Puebla, Puebla, Mexico }
\author{X.~Wang}
\author{Z.~Wang}
\affiliation{Department of Physics, Missouri University of Science and Technology, Rolla, MO, US}
\author{I.J.~Watson}
\email[Contact author: ]{ian.james.watson@cern.ch}
\affiliation{University of Seoul, Seoul, Republic of Korea}
\author{H.~Wu}
\affiliation{Dept. of Physics and Wisconsin IceCube Particle Astrophysics Center, University of Wisconsin{\textemdash}Madison, Madison, WI, USA}
\author{S.~Yu}
\affiliation{Department of Physics, Pennsylvania State University, University Park, PA, USA }
\author{S.~Yun-C\'{a}rcamo}
\affiliation{Department of Physics, Drexel University, Philadelphia, PA 19104, USA}
\author{X.~Zhang}
\affiliation{Institute of Nuclear Physics Polish Academy of Sciences, PL-31342 IFJ-PAN, Krakow, Poland }
\author{H.~Zhou}
\affiliation{Tsung-Dao Lee Institute \& School of Physics and Astronomy, Shanghai Jiao Tong University, 800 Dongchuan Rd, Shanghai, SH 200240, China}
\author{C.~de Le\'{o}n}
\affiliation{Universidad Michoacana de San Nicol\'{a}s de Hidalgo, Morelia, Mexico }
\collaboration{HAWC Collaboration}
\noaffiliation

\date{\today}

\begin{abstract}
Axion-like particles (ALPs) can mix with photons in astrophysical magnetic fields, modifying the propagation of very-high-energy gamma rays from distant blazars and producing an energy-dependent enhancement of the observed flux relative to the standard expectation from absorption on the extragalactic background light (EBL). We search for this signature through a joint likelihood analysis of seven TeV BL~Lac objects at redshifts $z = 0.14$--$0.60$, combining approximately 3,000~days of HAWC observations with published High Energy Spectroscopic System (H.E.S.S.), Major Atmospheric Gamma Imaging Cherenkov Telescopes (MAGIC), and Very Energetic Radiation Imaging Telescope Array System(VERITAS) spectra. We find no statistically significant ALP signal and derive exclusion limits using the $\mathrm{CL}_s$ method. We exclude photon-ALP couplings at the 95\% $\mathrm{CL}_s$ level down to $5.5 \times 10^{-12}~\mathrm{GeV}^{-1}$ for $m_a \lesssim 4$~neV and $4.6 \times 10^{-12}~\mathrm{GeV}^{-1}$ near $m_a \approx 6$--$9$~neV. The systematic uncertainty on the excluded parameter-space area is $-36.5\%$ to $+32.6\%$, dominated by the choice of EBL model. Because the survival probability oscillates with the coupling, the excluded region is not simply connected, and we quote at each mass the lowest excluded coupling rather than an upper bound. The combined limit is set almost entirely by the highest-redshift detected source, PG~1553+113.
\end{abstract}

\maketitle

\section{Introduction} \label{sec:intro}

Axion-like particles (ALPs) are pseudoscalar particles that arise in many extensions of the Standard Model~\citep{JaeckelRingwald2010,IrastorzaRedondo2018}, including string theory compactifications~\citep{Svrcek2006,Arvanitaki2010}. Unlike the quantum chromodynamics (QCD) axion, whose mass and photon coupling are both set by the scale of Peccei-Quinn symmetry breaking~\citep{PQAxion,Weinberg1978,Wilczek1978}, ALPs have an independent mass, $m_a$, and two-photon coupling, $g_{a\gamma\gamma}$. They are also viable candidates for cold dark matter~\citep{Preskill1983,Abbott1983,Dine1983,Marsh2016}. The two-photon coupling induces photon-ALP oscillations in external magnetic fields.

High-energy gamma rays emitted by distant extragalactic sources undergo energy-dependent attenuation via electron-positron pair production ($\gamma\gamma \rightarrow e^{+}e^{-}$) upon interacting with the extragalactic background light (EBL) \citep{BreitWheeler,GouldSchreder1967,Franceschini2008}. The EBL consists of the accumulated radiation from star formation and dust reprocessing over cosmic history. The optical depth, $\tau(E, z)$, quantifies this attenuation as a function of gamma-ray energy, $E$, and source redshift, $z$, and depends on both the cosmological distance and the evolution of the EBL. This results in an exponential suppression of the intrinsic flux, $\phi_{\mathrm{int}}(E)$:
\begin{equation}
    \phi_{\mathrm{obs}}(E) = \phi_{\mathrm{int}}(E) \, e^{-\tau(E,z)}.
    \label{eq:ebl}
\end{equation}
This attenuation becomes significant in the very-high-energy (VHE; $E \gtrsim 100$~GeV) regime and increases steeply with both energy and redshift.

Gamma rays can oscillate into ALPs when traversing magnetized environments. Because ALPs do not interact with the EBL, they propagate through intergalactic space unattenuated. If these ALPs subsequently reconvert into photons within other magnetized regions (e.g., the Galactic magnetic field; GMF), the net effect is an enhancement of the observed VHE gamma-ray flux relative to the standard EBL attenuation predictions. This phenomenon, referred to as ``TeV transparency''~\citep{DeAngelis2007,DeAngelis2011}, is most pronounced at energies where EBL absorption is otherwise severe, so VHE observations provide a sensitive probe of the ALP parameter space. Photon-ALP mixing also produces spectral irregularities, since the conversion probability oscillates with energy~\citep{HESS2013ALP,FermiLAT2016}. Searches for such irregularities require fine energy resolution over a well-measured spectrum, whereas the present work targets the transparency signature at multi-TeV energies, at which the attenuated flux is smallest and the relative enhancement largest. Several analyses have reported indications that the VHE gamma-ray opacity of the Universe is lower than predicted by EBL models~\citep{HornsMeyer2012,RubtsovTroitsky2014}, an effect that photon-ALP conversion has been invoked to explain~\citep{Meyer2013,DeAngelis2011}, although the significance of these indications has been contested~\citep{BiteauWilliams2015}. The present search constrains the ALP parameter space relevant to such interpretations.

While Imaging Atmospheric Cherenkov Telescopes (IACTs) are sensitive over a range from tens of GeV to tens of TeV~\citep{deNauroisMazin2015}, in practice their spectral measurements of distant blazars are limited to below a few TeV by increasing EBL attenuation~\citep{Dominguez2011}. The HAWC Observatory is sensitive from 0.3~TeV to 100~TeV~\citep{HAWCPass5} and provides coverage at multi-TeV energies where EBL attenuation is largest. In this work, we perform a joint likelihood fit combining published IACT flux points (described by a Gaussian likelihood) with HAWC data (described by a Poisson likelihood). The IACT flux points constrain the intrinsic blazar energy spectrum, while the HAWC data provide sensitivity to the ALP-induced transparency effect at higher energies. A previous search using the public 3HWC catalog fluxes set limits from a combined sample of seven TeV blazars~\citep{HAWC_ALP_2023}. The present work analyzes the same seven sources with the full HAWC dataset and its corresponding instrument response function (IRF).

HAWC is located at an altitude of 4,100~m on the Sierra Negra volcano in Puebla, Mexico~\citep{HAWC2023}. The main array comprises 300 water Cherenkov detectors (WCDs), each instrumented with four photomultiplier tubes (PMTs), covering an area of approximately 22,000~m$^2$. Since late 2018 the main array has been surrounded by an outrigger array of 345 smaller WCDs~\citep{OutriggerStatus, OutriggerReco}. The outriggers enlarge the instrumented area by approximately a factor of four and improve the reconstruction of showers with cores outside the main array above a few TeV. Outrigger data are not used in the Pass~5 reconstruction employed here. HAWC observes two-thirds of the sky every day with a duty cycle above 93\%, enabling long-term monitoring of VHE gamma-ray sources~\citep{HAWC2023}.

\section{Photon-ALP Oscillations} \label{sec:theory}

The ALP-photon interaction is described by the effective Lagrangian:
\begin{equation}
    \mathcal{L}_{a\gamma\gamma} = -\frac{1}{4} g_{a\gamma\gamma} a F_{\mu\nu} \tilde{F}^{\mu\nu} = g_{a\gamma\gamma} a \, \mathbf{E} \cdot \mathbf{B},
\end{equation}
where $a$ is the ALP field, $F_{\mu\nu}$ is the electromagnetic field strength tensor, and $\tilde{F}^{\mu\nu}$ is its dual. This interaction allows photon-ALP conversion in external magnetic fields.

The photon-ALP system is described by a state vector, $\Psi = (A_x, A_y, a)^T$, where $A_x$ and $A_y$ are the orthogonal photon polarization amplitudes, and $a$ is the ALP amplitude. The propagation of a monochromatic photon beam of energy $E$ through a cold plasma embedded in a homogeneous magnetic field is governed by the equation of motion~\citep{Raffelt1988,gammaALPs}:
\begin{equation}
    \left( i\frac{d}{d\zeta} + E + \mathcal{M}_0 \right) \Psi(\zeta) = 0.
\end{equation}
Here $\zeta$ denotes the proper distance along the line of sight, measured from the emission region towards the observer. The beam propagates along the $\zeta$ axis, and the $x$ and $y$ axes span the transverse plane. The symbol $z$ is reserved for redshift.
Neglecting Faraday rotation, which scales as $E^{-2}$ and is negligible at VHE energies~\citep{Raffelt1988}, the mixing matrix $\mathcal{M}_{0}$ is given by:
\begin{equation}
    \mathcal{M}_0 = \begin{pmatrix}
        \Delta_\perp & 0 & 0 \\
        0 & \Delta_\parallel & \Delta_{a\gamma} \\
        0 & \Delta_{a\gamma} & \Delta_a
    \end{pmatrix}.
\end{equation}
Here the $y$~axis is chosen along $\mathbf{B}_T$, so that $\Delta_\perp$ and $\Delta_\parallel$ correspond to the photon polarization states perpendicular and parallel to the transverse magnetic field, respectively. These terms include plasma effects and quantum electrodynamics (QED) vacuum polarization:
\begin{equation}
    \Delta_\perp = \Delta_{\mathrm{pl}} + 2\Delta_{\mathrm{QED}}, \quad \Delta_\parallel = \Delta_{\mathrm{pl}} + \frac{7}{2}\Delta_{\mathrm{QED}},
\end{equation}
where $\Delta_{\mathrm{pl}} = -\omega_{\mathrm{pl}}^2/(2E)$ is the plasma term with $\omega_{\mathrm{pl}}$ being the plasma frequency, and $\Delta_{\mathrm{QED}} = (\alpha E / 45\pi)(B_T/B_{\mathrm{cr}})^2$ represents the QED vacuum polarization term with the critical magnetic field $B_{\mathrm{cr}} \approx 4.4 \times 10^{13}~\mathrm{G}$~\citep{HeisenbergEuler1936,Schwinger1951}. The ALP kinetic term is defined as $\Delta_a = -m_a^2/(2E)$, and the photon-ALP mixing term is $\Delta_{a\gamma} = g_{a\gamma\gamma} B_T / 2$.

These equations of motion are solved using the transfer matrix formalism~\citep{DeAngelis2011}. For an initially unpolarized photon beam, the initial density matrix is set to $\rho(0)=\mathrm{diag}(1/2,1/2,0)$, representing equal probabilities for the two orthogonal photon polarization states and a zero initial ALP abundance. The density matrix $\rho(\zeta) = \langle \Psi(\zeta)\Psi^\dagger(\zeta) \rangle$, where $\langle \cdot \rangle$ denotes the ensemble average over the polarization states of the beam, then evolves along the line of sight according to: 
\begin{equation}
    \rho(\zeta) = T(\zeta, 0; E) \, \rho(0) \, T^\dagger(\zeta, 0; E).
\end{equation}
Here, $T$ denotes the transfer matrix satisfying $\Psi(\zeta) = T(\zeta, 0; E)\Psi(0)$. For a line of sight traversing $N$ distinct domains with varying magnetic field configurations, delimited by $\zeta_{0}<\zeta_{1}<\dots<\zeta_{N}$, the total transfer matrix is given by the ordered product:
\begin{equation}
    T(\zeta_N, \zeta_0; E) = \prod_{i=0}^{N-1} T(\zeta_{i+1}, \zeta_i; E).
\end{equation}
To evaluate the observable photon survival probability, $P_{\gamma\gamma}$, we project the final density matrix onto the observable photon states and take the trace:
\begin{equation}
    P_{\gamma\gamma} = \mathrm{Tr}\left[ (\rho_{11} + \rho_{22}) \, T \, \rho(0) \, T^\dagger \right],
\end{equation}
where $\rho_{11} = \mathrm{diag}(1,0,0)$ and $\rho_{22} = \mathrm{diag}(0,1,0)$ project onto the two orthogonal photon polarization states.

Absorption on the EBL enters the same ordered product as an additional, non-unitary transfer matrix acting only on the photon components,
\begin{equation}
    T_{\mathrm{EBL}} = \mathrm{diag}\left(e^{-\tau(E,z)/2},\; e^{-\tau(E,z)/2},\; 1\right),
\end{equation}
inserted between the environments local to the source and the Galactic field, since ALPs cross the intergalactic medium unattenuated. For $g_{a\gamma\gamma} = 0$ the photon and ALP sectors decouple and the product reduces to $P_{\gamma\gamma} = e^{-\tau(E,z)}$, recovering Eq.~\eqref{eq:ebl}.

Efficient photon-ALP mixing sets in above a critical energy threshold, $E_{\mathrm{crit}}$, beyond which the mixing angle $\theta$, defined by $\tan 2\theta = 2\Delta_{a\gamma}/(\Delta_\parallel - \Delta_a)$, approaches its maximal value of $\pi/4$. The threshold lies where twice the off-diagonal mixing term, $2\Delta_{a\gamma} = g_{a\gamma\gamma} B_T$, equals the kinematic phase difference between the photon and ALP states, $|\Delta_a - \Delta_\parallel|$:
\begin{equation}
    E_{\mathrm{crit}} = \frac{|m_a^2 - \omega_{\mathrm{pl}}^2|}{2 g_{a\gamma\gamma} B_T}.
\end{equation}
In the high-energy limit ($E \gg E_{\mathrm{crit}}$), the system enters the strong mixing regime where the amplitude of photon-ALP oscillations is maximal. Conversely, for $E \ll E_{\mathrm{crit}}$, the mixing is strongly suppressed by the mass and plasma terms. Along the line of sight, these oscillations can occur in several magnetized environments. In the source vicinity, the blazar jet and, where present, the turbulent intra-cluster magnetic field (ICMF) act as the primary conversion regions. The $\mu$G-scale GMF of the Milky Way is the final conversion region before detection. 
Following \citep{HAWC_ALP_2023}, we neglect the intergalactic magnetic field. The present-day primordial field, integrated over all scales, is constrained by the effect of magnetically induced baryon inhomogeneities on cosmic recombination to $47$~pG for a scale-invariant spectrum and $8.9$~pG for a Batchelor spectrum at 95\% confidence~\citep{JedamzikSaveliev2019}. We evaluated the impact of the omission for PG~1553+113, which dominates the combined limit (Section~\ref{sec:contrib}), at a fixed turbulent-field realization. An intergalactic component at the less restrictive bound with a 1~Mpc coherence length changes $P_{\gamma\gamma}$ across the HAWC band by a median of $0.3\%$. Assigning the full bound to a single coherent domain maximizes the conversion, so this value is an upper estimate.

\section{Sources and Line-of-Sight Modeling}

\subsection{Source Selection}

Parameter values are taken from the primary literature and are listed with their sources in Tables~\ref{tab:sources_coords},~\ref{tab:bfield_params} and~\ref{tab:sources}; departures from Ref.~\citep{HAWC_ALP_2023} are noted in the text where they occur.

The initial sample comprises the nine blazars considered in \citep{HAWC_ALP_2023}. We exclude two of them, PKS~1222+216 and MAGIC~J2001+439, as in \citep{HAWC_ALP_2023}, so that the samples are identical and the results directly comparable. The criterion in both analyses is that a single steady-state spectrum describes each source over the full exposure. The line-of-sight model predicts a time-independent $P_{\gamma\gamma}(E)$ acting on one intrinsic spectrum, and the time-averaged spectrum of a source varying by orders of magnitude does not represent such a quantity. The flat-spectrum radio quasar PKS~1222+216 shows flux-doubling timescales of $\sim$10~minutes during major outbursts~\citep{Tavecchio2011}, and MAGIC~J2001+439 was detected at VHE only during a single flaring episode in 2010 July~\citep{Aleksic2014_J2001}.

\begin{table*}[ht!]
\caption{Source coordinates and redshifts. Coordinates are J2000. Redshifts marked $^a$ are lower limits (VER~J0521+211, PKS~1424+240) or the lower bound of the determination (PG~1553+113), adopted as the analysis values (see text). The IACT column indicates the observatory providing the spectral flux points used in the joint likelihood analysis. $\sqrt{\mathrm{TS}_{\mathrm{HAWC}}}$ is the HAWC-only significance of the no-ALP model, with $\mathrm{TS}_{\mathrm{HAWC}} = 2[\ln\mathcal{L}_{\mathrm{HAWC}}(\hat{K}) - \ln\mathcal{L}_{\mathrm{HAWC}}(K = 0)]$ obtained by profiling the flux normalization with the spectral index fixed to the joint no-ALP best fit (Section~\ref{sec:contrib}). For 1ES~1011+496 and 3C~66A the profiled normalization is negative and $\mathrm{TS}_{\mathrm{HAWC}} < 0$ within the tolerance of the fit. The signed value $-\sqrt{|\mathrm{TS}_{\mathrm{HAWC}}|}$ is quoted.}
\label{tab:sources_coords}
\begin{ruledtabular}
\begin{tabular}{lccccc}
Source & R.A. & Dec. & $z$ & IACT & $\sqrt{\mathrm{TS}}_{\mathrm{HAWC}}$ \\
 & (deg) & (deg) & & & \\
\hline
1ES~0229+200 & 38.222 & 20.273 & 0.140 & H.E.S.S. \citep{1ES0229} & 0.6 \\
1ES~1011+496 & 153.767 & 49.434 & 0.212 & MAGIC \citep{1ES1011} & $-0.01$ \\
1ES~1218+304 & 185.360 & 30.191 & 0.182 & VERITAS \citep{1ES1218} & 6.6 \\
3C~66A & 35.673 & 43.043 & 0.34 & MAGIC \citep{3C66A} & $-0.10$ \\
PKS~1424+240 & 216.750 & 23.794 & 0.60$^a$ & VERITAS \citep{PKS1424} & 2.6 \\
VER~J0521+211 & 80.438 & 21.214 & 0.18$^a$ & VERITAS \citep{VERJ0521} & 5.9 \\
PG~1553+113 & 238.929 & 11.190 & 0.491$^a$ & VERITAS \citep{PG1553} & 5.7 \\
\end{tabular}
\end{ruledtabular}
\end{table*}

The remaining seven blazars are analyzed in this work. Table~\ref{tab:sources_coords} lists their coordinates, redshifts, and the IACT whose published spectral measurements are used in the joint likelihood fit~\citep{1ES0229,1ES1011,1ES1218,3C66A,PKS1424,VERJ0521,PG1553}. The sample spans redshifts from $z = 0.14$ (1ES~0229+200) to $z = 0.60$ (PKS~1424+240), covering a range of EBL optical depths. For VER~J0521+211 and PKS~1424+240, only redshift lower limits are available~\citep{Paiano2017,Furniss2013}, and we adopt these bounds. This choice is conservative, since a lower redshift reduces the EBL attenuation and hence the predicted ALP-induced enhancement. For 3C~66A we adopt $z = 0.34$, from the identification of its host galaxy cluster~\citep{TorresZafra2018}. For PG~1553+113, we adopt $z = 0.491$, the lower bound of the determination by \citep{Sahu2025} under the baseline EBL model of this analysis. 
The IACT measurements constrain the intrinsic spectra below a few TeV, where EBL attenuation is smaller. For PG~1553+113, which dominates the combined limit (Section~\ref{sec:contrib}), the baseline EBL model~\citep{Dominguez2011} gives $e^{-\tau} = 0.42$ at 200~GeV and $2.6 \times 10^{-2}$ at 500~GeV, compared with $4 \times 10^{-7}$ at 5~TeV. The IACT data thus fix the intrinsic spectrum where the EBL correction is small, while HAWC covers the multi-TeV range where the ALP-induced enhancement is largest.

Four sources (1ES~0229+200, 3C~66A, PKS~1424+240, and PG~1553+113) are associated with galaxy clusters or groups and are given an ICMF component. For 1ES~0229+200, PKS~1424+240 and PG~1553+113 the association follows the cross-correlation of \citep{MeyerConrad2014} against the GMBCG~\citep{GMBCG2010} and WHL~\citep{WHL2012} cluster catalogs, which required the blazar to lie within the cluster radius and within $|z - z_{\mathrm{cl}}| \leq 0.05$ in redshift. That for 3C~66A follows \citep{HAWC_ALP_2023}. The ICM field of 1ES~0229+200 is normalized to the value measured for the 3C~449 group~\citep{Guidetti2010}, following \citep{HAWC_ALP_2023}.  
For the other three sources, whose environments are not individually characterized, the normalization is fixed to $B_0^{\mathrm{ICM}} = 1.0~\mu$G, the value adopted in \citep{MeyerConrad2014} and retained in \citep{HAWC_ALP_2023}. Central field strengths inferred from Faraday rotation measures in groups and poor clusters are typically a few $\mu$G~\citep{Guidetti2010}; the adopted value lies below this range. Because the conversion probability increases with $B_0^{\mathrm{ICM}}$, the choice is conservative.
The GMF is included for all sources. Following \citep{HAWC_ALP_2023}, 1ES~0229+200, for which no jet-field determination from spectral energy distribution (SED) is available (Table~\ref{tab:sources}), is treated without a jet component, with an enhanced cluster field ($B_0^{\mathrm{ICM}} = 3.5~\mu$G) as the only conversion region near the source. The blazar jet magnetic field parameters are taken from multi-wavelength SED modeling in the literature. The adopted values are listed in Table~\ref{tab:sources}.

\subsection{Galactic Magnetic Field}

For the GMF, we adopt the model of \citep{JanssonFarrar2012} with its original best-fit parameters (hereafter Jansson12a). To evaluate model-dependent uncertainties, we also consider the two updated parameter sets obtained from fits to Planck dust polarization data (Jansson12b, Jansson12c)~\citep{Planck2016GMF}. Only the regular field components of the model enter the calculation. The striated and small-scale random components are not included. 
The transverse field strength $B_{T}$ along the line of sight varies significantly between sources owing to their different Galactic coordinates. The spread among the three parameter sets is propagated as the systematic uncertainty of the Jansson--Farrar model (Section~\ref{sec:syst}).

\subsection{Extragalactic Background Light} \label{sec:ebl}

We adopt the model of \citep{Dominguez2011} as the baseline for EBL attenuation, as in \citep{HAWC_ALP_2023}, which allows a direct comparison with that search. To evaluate systematic uncertainties, we consider alternative models by \citep{Franceschini2008}, \citep{Gilmore2012}, \citep{Finke2022}, and \citep{SaldanaLopez2021}. The optical depth $\tau(E, z)$ differs among these models. The attenuation factor $e^{-\tau}$ falls steeply with both energy and redshift at TeV energies, and the high-redshift sources are therefore the most sensitive to ALP-induced transparency (Section~\ref{sec:contrib}), while the EBL model dominates the systematic budget (Section~\ref{sec:syst}).

\subsection{Source Magnetic Fields}

\begin{table}[ht!]
\caption{Default magnetic field model parameters. $^a$1ES~0229+200 uses $B_0^{\mathrm{ICM}} = 3.5~\mu$G. Source-specific jet parameters ($B_0^{\mathrm{jet}}$, $n_0^{\mathrm{jet}}$, $\delta$) from SED modeling are in Table~\ref{tab:sources}. Emission region $r_0 = R_{\mathrm{VHE}}/\tan(1/\Gamma_{\mathrm{L}})$ with $R_{\mathrm{VHE}} \sim 10^{16}$~cm and $\Gamma_{\mathrm{L}}$ the bulk Lorentz factor. Parameter names follow \texttt{gammaALPs}. Here $r_{\mathrm{Abell}}$ is the radial extent of the ICM field region. The ICMF turbulence parameters ($k_L$, $k_H$, $q$) and the density-profile parameters ($n_0^{\mathrm{ICM}}$, $r_{\mathrm{core}}$, $\beta_{\mathrm{ICM}}$) are the values measured for the 3C~449 group~\citep{Guidetti2010} in the form adopted in \citep{MeyerConrad2014}; the cluster extent $r_{\mathrm{Abell}}$, the jet indices $\alpha$ and $\beta_{\mathrm{jet}}$, and the jet extent $R_{\mathrm{jet}}$ follow \citep{Meyer2014} and \citep{HAWC_ALP_2023}. The derivation of $q = -2.53$ is given in Section~\ref{sec:icmf}.}
\label{tab:bfield_params}
\begin{ruledtabular}
\begin{tabular}{lcc}
Parameter & Value & Description \\
\hline
\multicolumn{3}{c}{\textit{ICMF (Gaussian Turbulence)}} \\
\hline
$B_0^{\mathrm{ICM}}$ & 1.0~$\mu$G & Cluster central field strength$^a$ \\
$k_L$ & 0.015~kpc$^{-1}$ & Low-$k$ cutoff \\
$k_H$ & 5.0~kpc$^{-1}$ & High-$k$ cutoff \\
$q$ & $-2.53$ & Spectral index \\
$n_0^{\mathrm{ICM}}$ & $3.7 \times 10^{-3}$~cm$^{-3}$ & Central $e^-$ density \\
$r_{\mathrm{core}}$ & 19.33~kpc & Core radius \\
$\beta_{\mathrm{ICM}}$ & 0.42 & Beta model index \\
$r_{\mathrm{Abell}}$ & 100~kpc & ICM field extent \\
$N_{\mathrm{sim}}$ & 1000 & Realizations \\
\hline
\multicolumn{3}{c}{\textit{Jet (Toroidal Field)}} \\
\hline
$\alpha$ & $-1$ & $B$-field radial index \\
$\beta_{\mathrm{jet}}$ & $-2$ & $n_e$ radial index \\
$R_{\mathrm{jet}}$ & 1000~pc & Max jet extent \\
Helical & No & Toroidal only \\
\end{tabular}
\end{ruledtabular}
\end{table}

\begin{table}[ht!]
\caption{Magnetic field parameters for each source. $B_0^{\mathrm{jet}}$: jet magnetic field strength at the VHE emission region; $n_0^{\mathrm{jet}}$: jet electron density at the VHE emission region; $\delta$: Doppler factor; $B_0^{\mathrm{ICM}}$: cluster central field strength of the turbulent ICMF. The last column gives the source of the field parameters: the SED model for the jet parameters, or, for 1ES~0229+200, the rotation-measure study for the cluster field.} Sources without ICMF use the Jet+GMF scenario; sources with ICMF use Jet+ICMF+GMF (except 1ES~0229+200, which uses ICMF+GMF only).
\label{tab:sources}
\begin{ruledtabular}
\begin{tabular}{lccccc}
Source & $B_0^{\mathrm{jet}}$ (G) & $n_0^{\mathrm{jet}}$ (cm$^{-3}$) & $\delta$ & $B_0^{\mathrm{ICM}}$ ($\mu$G) & Ref. \\
\hline
1ES~0229+200 & --- & --- & --- & 3.5 & \citep{Guidetti2010} \\
1ES~1011+496 & 0.048 & $7.0 \times 10^{2}$ & 26 & --- & \citep{1ES1011} \\
1ES~1218+304 & 0.22 & $1.0 \times 10^{3}$ & 26 & --- & \citep{1ES1218sed} \\
3C~66A & 0.21 & $1.0 \times 10^{4}$ & 40 & 1.0 & \citep{3C66Ased} \\
PKS~1424+240 & 0.033 & $3.1 \times 10^{3}$ & 30 & 1.0 & \citep{PKS1424sed} \\
VER~J0521+211 & 0.1 & $3.15 \times 10^{4}$ & 12 & --- & \citep{VERJ0521sed} \\
PG~1553+113 & 0.5 & $5.35 \times 10^{3}$ & 35 & 1.0 & \citep{PG1553sed} \\
\end{tabular}
\end{ruledtabular}
\end{table}

We consider two types of magnetic field environments near the blazars, implemented using the \texttt{gammaALPs} package~\citep{gammaALPs}:

\subsubsection{Blazar Jet Magnetic Field}

The jet magnetic field is modeled as toroidal. The toroidal component falls as $r^{-1}$ and dominates at the distances from the central engine where the conversion occurs~\citep{Meyer2014}. Only the component transverse to the line of sight contributes to the mixing, and at the small viewing angles of BL~Lac objects the toroidal field is nearly transverse. The same prescription is adopted in \citep{HAWC_ALP_2023}. The radial dependence is:
\begin{equation}
    B(r) = B_0^{\mathrm{jet}} \left(\frac{r}{r_0}\right)^{\alpha},
\end{equation}
where $B_0^{\mathrm{jet}}$ is the jet magnetic field strength at the reference radius $r_0$, and $\alpha = -1$ is the power-law index. The VHE emission region is located at $r_0 \approx R_{\mathrm{VHE}}/\tan(1/\Gamma_{\mathrm{L}})$, where $R_{\mathrm{VHE}} \sim 10^{16}$~cm is the typical emission-region size in the SED models cited in Table~\ref{tab:sources} ($R = (1$--$3.3) \times 10^{16}$~cm) and $\Gamma_{\mathrm{L}}$ is the bulk Lorentz factor. The electron density in the jet follows a similar power-law profile:
\begin{equation}
    n_e(r) = n_0^{\mathrm{jet}} \left(\frac{r}{r_0}\right)^{\beta_{\mathrm{jet}}},
\end{equation}
with $\beta_{\mathrm{jet}} = -2$. The jet extends to a maximum radius of $R_{\mathrm{jet}} = 1000$~pc (Table~\ref{tab:bfield_params}). The adopted jet magnetic field strengths, $B_0^{\mathrm{jet}}$, range from 0.033~G (PKS~1424+240) to 0.5~G (PG~1553+113), as detailed in Table~\ref{tab:sources}. The Doppler factor $\delta$ of each source (Table~\ref{tab:sources}) sets the transformation of the observed photon energy into the jet comoving frame, in which the mixing is evaluated. The $\alpha = -1$ power-law index is adopted from the literature~\citep{Meyer2014} and is consistent with a toroidal field geometry in a conical jet.

The source-specific jet parameters ($B_0^{\mathrm{jet}}$, $n_0^{\mathrm{jet}}$, $\delta$) are taken from the SED modeling of each source~\citep{1ES1011,1ES1218sed,3C66Ased,PKS1424sed,VERJ0521sed,PG1553sed}, using the VHE-emitting zone in the case of two-zone models. For VER~J0521+211 and PKS~1424+240, we adopt the electron densities reported in the original SED analyses, $3.15 \times 10^{4}~\mathrm{cm}^{-3}$ and $3.1 \times 10^{3}~\mathrm{cm}^{-3}$~\citep{VERJ0521sed,PKS1424sed}, which differ from the values tabulated in \citep{HAWC_ALP_2023}. For 1ES~1218+304 and 3C~66A, whose SED analyses do not directly report an electron number density, we adopt the values used in \citep{HAWC_ALP_2023}.

\subsubsection{Intra-Cluster Magnetic Field} \label{sec:icmf}

For sources located in galaxy cluster environments, the ICMF is modeled as a stochastic Gaussian random field following the formalism of \citep{Meyer2014}. The magnetic field turbulence is characterized by a power spectrum:
\begin{equation}
    M(k) \propto k^{q}, \quad k_L \leq k \leq k_H,
    \label{eq:turbspec}
\end{equation}
where $k_L$ and $k_H$ represent the lower and upper wavenumber cutoffs, determining the maximum and minimum scales of the turbulence, respectively. We adopt a spectral index of $q = -2.53$, flatter than the Kolmogorov value of $-11/3$~\citep{Kolmogorov1941}. This index and the turbulence cutoffs of Table~\ref{tab:bfield_params} are measured rather than assumed. The measurement is of the Fanaroff-Riley~I radio galaxy 3C~449, which lies in a group of galaxies comparable to the environments considered here. Rotation-measure and depolarization data for that source give a broken power-law magnetic power spectrum~\citep{Guidetti2010}. That work writes the spectrum as $\propto f^{-n}$, with the opposite sign convention for the exponent to Eq.~\eqref{eq:turbspec}, so that Kolmogorov turbulence corresponds to $n = 11/3$. Its fitted slopes are $n = 2.98$ and $2.07$ at scales smaller and larger than a break at $\approx 11$~kpc, with fluctuation scales bounded by $\approx 0.2$ and $\approx 65$~kpc and a central field strength of $3.5 \pm 1.2~\mu$G. \citep{MeyerConrad2014} approximated the broken power law by the average of the two slopes, $n = 2.53$, which is $q = -2.53$ in the convention of Eq.~\eqref{eq:turbspec}. They took the measured outer and inner scales as the cutoffs $k_L = 0.015$~kpc$^{-1}$ and $k_H = 5$~kpc$^{-1}$. The same measurement supplies the $3.5~\mu$G cluster field adopted for 1ES~0229+200 (Table~\ref{tab:sources}). This prescription was retained by \citep{HAWC_ALP_2023} and is retained here. Rotation measures of poor environments generally yield turbulence spectra flatter than Kolmogorov~\citep{Guidetti2011}, and a flatter spectrum reduces the ALP conversion probability~\citep{Meyer2014}, so the choice is conservative. The transverse components of the turbulent magnetic field, which are directly responsible for the photon-ALP mixing, are computed by generating random spatial realizations from this power spectrum. The field strength is normalized to the cluster central value $B_0^{\mathrm{ICM}}$ (Table~\ref{tab:bfield_params}). Unlike $B_0^{\mathrm{jet}}$, which is anchored to source-specific SED modeling, this normalization is a generic cluster-environment assumption.

The cluster electron density, which governs the effective plasma mass, follows a standard beta model profile~\citep{Cavaliere1976}:
\begin{equation}
    n_e(r) = n_0^{\mathrm{ICM}} \left(1 + \frac{r^2}{r_{\mathrm{core}}^2}\right)^{-3\beta_{\mathrm{ICM}}/2},
\end{equation}
where $n_0^{\mathrm{ICM}}$ is the central electron density, $r_{\mathrm{core}}$ is the core radius, and $\beta_{\mathrm{ICM}}$ dictates the steepness of the density decline in the outer cluster regions. The density-profile parameters of Table~\ref{tab:bfield_params} are those measured for the 3C~449 environment~\citep{Guidetti2010}. The cluster extent follows \citep{HAWC_ALP_2023}.

To account for the stochastic nature of the turbulent ICMF, we generate $N_{\mathrm{sim}} = 1000$ independent magnetic field realizations for each source. The photon survival probabilities are evaluated at the 5\%, 50\%, and 95\% quantiles of the resulting distribution to quantify the associated systematic uncertainty. The ensemble size is set by the convergence of the median, which serves as the nominal survival probability (Section~\ref{sec:pgg}). With $N_{\mathrm{sim}} = 1000$, bootstrap resampling of the realizations yields a standard error on the median $P_{\gamma\gamma}$ below $2.7\%$. This holds for all four sources with an ICMF component over the full $(m_a, g_{a\gamma\gamma})$ grid and energy range. Default ICMF and jet parameters are listed in Table~\ref{tab:bfield_params}. The magnetic field scenario applied to each source (e.g., whether a jet or ICMF component is included) and the source-specific parameters are listed in Table~\ref{tab:sources}.

\subsection{Photon Survival Probability and ALP Parameter Space} \label{sec:pgg}

The photon survival probability, $P_{\gamma\gamma}(E; m_a, g_{a\gamma\gamma})$, is pre-computed with the \texttt{gammaALPs} package~\citep{gammaALPs} on a $200 \times 200$ logarithmic grid and interpolated during the fit. The grid spans ALP masses $m_a \in [0.1, 10^4]$~neV and photon-ALP couplings $g_{a\gamma\gamma} \in [10^{-13}, 10^{-8}]$~GeV$^{-1}$. The survival probability is tabulated at 100 logarithmically spaced energies from 1~GeV to 1~PeV. The spacing in all three variables is chosen for the accuracy of the interpolation, as $P_{\gamma\gamma}$ oscillates in both coupling and energy. The interpolated $P_{\gamma\gamma}$ multiplies the intrinsic spectrum, and the product is forward-folded through the HAWC instrument response. The detector energy resolution is thus applied after the ALP modulation.
For sources with stochastic ICMF components, no single turbulent-field realization is preferred a~priori. We therefore adopt the median (50\% quantile) of the realization distribution, which is insensitive to its extremes, as the nominal survival probability and propagate the spread between the 5\% and 95\% quantiles as a systematic uncertainty (Section~\ref{sec:syst}). This treatment differs from the mean survival probability adopted by \citep{HAWC_ALP_2023}. Other prescriptions in the literature include the likelihood-quantile method of \citep{FermiLAT2016} and the ensemble-coverage criterion of \citep{HAWC_M87_2026}.

\section{HAWC Data Analysis} \label{sec:analysis}

\subsection{Dataset}

This analysis uses the HAWC ``Pass 5'' dataset~\citep{HAWCPass5, HAWC:2025nyi}, reprocessed with the improved reconstruction. The data were taken between 2014 November and 2024 January and amount to approximately 3,000~days of livetime after data-quality selection.
The dataset employs a multilayer-perceptron-based gamma-hadron separation~\citep{HAWC:2025nyi}, which improves sensitivity in the low fractional-hit bins relative to traditional cut-based methods~\citep{HAWCPass5}. This improvement is relevant for the soft, EBL-attenuated spectra of distant blazars. We include all available analysis bins, spanning fractional-hit bins 0 through 10 for both the on-array and off-array configurations, which together cover reconstructed energies from 0.3~TeV to beyond 100~TeV~\citep{HAWC:2025nyi}. Off-array events, whose shower cores land outside the main array, are reconstructed with poorer angular resolution than on-array events but add sensitivity, and are binned separately for that reason~\citep{HAWC:2025nyi}. 

\subsection{Spectral Model}

The intrinsic blazar spectrum is modeled as a simple power law:
\begin{equation}
    \phi_{\mathrm{int}}(E) = K \left(\frac{E}{E_0}\right)^{\Gamma},
\end{equation}
where $K$ is the differential flux normalization at the pivot energy $E_0$, and $\Gamma$ is the spectral index ($\Gamma < 0$ in this convention). A pivot energy $E_0 = 1$~TeV is used for all sources.  
During the likelihood optimization, $K$ and $\Gamma$ are treated as nuisance parameters and are profiled to determine their best-fit values at each ALP parameter point. The values obtained under the no-ALP hypothesis are listed in Table~\ref{tab:bestfit}. The power-law form is consistent with the published IACT measurements for all seven sources, which constrain it below a few TeV. 

\begin{table}[ht!]
\caption{Best-fit intrinsic spectral parameters of the no-ALP hypothesis, from the joint IACT-HAWC fit. $K$ and $\Gamma$ refer to $\phi_{\mathrm{int}}$, i.e.\ before EBL attenuation, at the common pivot energy $E_0 = 1$~TeV. These are the parameters of the null model against which the ALP hypothesis is tested at every grid point. Uncertainties are the $1\sigma$ profile-likelihood errors from the joint fit, with the ALP parameters fixed to zero.}
\label{tab:bestfit}
\begin{ruledtabular}
\begin{tabular}{lcc}
Source & $K$ ($10^{-12}$~TeV$^{-1}$cm$^{-2}$s$^{-1}$) & $\Gamma$ \\
\hline
1ES~0229+200 & $1.71^{+0.28}_{-0.32}$ & $-2.56^{+0.25}_{-0.29}$ \\
1ES~1011+496 & $4.88^{+1.95}_{-1.64}$ & $-2.31^{+0.27}_{-0.29}$ \\
1ES~1218+304 & $7.22^{+0.53}_{-0.54}$ & $-2.13^{+0.06}_{-0.06}$ \\
3C~66A & $1.26^{+1.48}_{-0.54}$ & $-2.75^{+0.39}_{-0.36}$ \\
PKS~1424+240 & $0.52^{+0.55}_{-0.19}$ & $-3.50^{+0.37}_{-0.21}$ \\
VER~J0521+211 & $3.72^{+0.65}_{-0.64}$ & $-2.65^{+0.14}_{-0.16}$ \\
PG~1553+113 & $19.91^{+2.60}_{-2.29}$ & $-2.17^{+0.08}_{-0.08}$ \\
\end{tabular}
\end{ruledtabular}
\end{table}

Above a few TeV the power law is an extrapolation, and it is there that it sets the no-ALP expectation against which the ALP hypothesis is tested. We adopt it as the baseline because VHE observations indicate that most blazars show no intrinsic spectral cutoff at $\sim$1~TeV~\citep{BiteauWilliams2015,HAWC_ALP_2023}. The choice is also structural to the method. Only the product $\phi_{\mathrm{int}}(E)\,P_{\gamma\gamma}(E)$ is observable, and in a transparency search the ALP signature is a smooth excess of multi-TeV flux over the extrapolated intrinsic spectrum. 

The observed flux at Earth is the product of the intrinsic spectrum and the photon survival probability:
\begin{equation}
    \phi_{\mathrm{obs}}(E) = \phi_{\mathrm{int}}(E) \times P_{\gamma\gamma}(E; m_a, g_{a\gamma\gamma}),
\end{equation}
where $P_{\gamma\gamma}(E; m_a, g_{a\gamma\gamma})$ is the energy-dependent photon survival probability derived from the transfer matrix formalism (Section~\ref{sec:theory}), which accounts for both EBL absorption and photon-ALP oscillations across all line-of-sight magnetic field environments.

\subsection{Joint Likelihood Analysis}

We perform a joint likelihood analysis combining IACT spectral flux points with the full HAWC dataset. The two instruments cover complementary energy ranges. IACTs reach hundreds of GeV to a few TeV, where EBL attenuation is smaller, and HAWC covers 0.3~TeV to 100~TeV, where it dominates.

This approach extends the methodology of \citep{HAWC_ALP_2023}, which used the public 3HWC catalog (a best-fit flux and a $2\sigma$ upper limit at 7~TeV per source) with a Gaussian likelihood approximation and excluded $g_{a\gamma\gamma} \gtrsim 3 \times 10^{-11}$~GeV$^{-1}$ for $m_a \lesssim 300$~neV. We instead use the full HAWC Poisson likelihood with the complete IRFs. The limit of \citep{HAWC_ALP_2023} was set at the $3\sigma$ level by applying Wilks' theorem~\citep{Wilks1938} to the test statistic. The present limits are calibrated with pseudo-experiments and quoted at 95\% CL$_s$~\citep{Junk1999,Read2002}, so the two are not defined identically. The CL$_s$ construction is standard in collider searches; among gamma-ray ALP searches it has been applied to Mrk~421~\citep{PhysRevD.109.063003}, which likewise generates pseudo-data under the ALP hypothesis.

Assuming the IACT and HAWC measurements to be statistically independent, and the seven sources independent of one another, the combined log-likelihood is:
\begin{equation}
    \ln \mathcal{L}_{\mathrm{total}} = \sum_{s} \left[ \ln \mathcal{L}_{\mathrm{IACT},s} + \ln \mathcal{L}_{\mathrm{HAWC},s} \right],
\end{equation}
where $s$ runs over the seven blazars. The ALP parameters $(m_a, g_{a\gamma\gamma})$ are common to all sources, while each source carries its own intrinsic-spectrum nuisance parameters $(K_s, \Gamma_s)$, profiled independently. The combined test statistic at a grid point is therefore the sum of the per-source test statistics, $q_{\mathrm{combined}} = \sum_s q_s$. The independence applies to the measurements. The ALP predictions for the seven sources remain correlated through the common EBL and GMF models, whose systematic variation is treated in Section~\ref{sec:syst}.

\subsubsection{IACT Likelihood}

For the IACT data, we use published spectral flux points listed in Table~\ref{tab:sources_coords}, from H.E.S.S., MAGIC, and VERITAS observations of the seven sources~\citep{1ES0229,1ES1011,1ES1218,3C66A,PKS1424,VERJ0521,PG1553}. To account for the finite energy resolution of the IACTs, the predicted flux in each reconstructed-energy bin $i$ (width $\Delta E_i$) is obtained by folding the model with the energy dispersion $D(E_{\mathrm{rec}} \,|\, E_{\mathrm{true}})$. The dispersion is taken to be Gaussian in $E_{\mathrm{rec}}$, with mean $E_{\mathrm{true}}$ and standard deviation $\sigma_{E}=0.1\,E_{\mathrm{true}}$, corresponding to a nominal energy resolution of 10\%, the approximation adopted in \citep{HAWC_ALP_2023}:
\begin{equation}
    \tilde{\phi}(E_{\mathrm{rec}}) = \int dE_{\mathrm{true}} \, D(E_{\mathrm{rec}} \,|\, E_{\mathrm{true}}) \, \phi_{\mathrm{obs}}(E_{\mathrm{true}}),
\end{equation}
\begin{equation}
    \phi_i^{\mathrm{pred}} = \frac{1}{\Delta E_i} \int_{\Delta E_i} dE_{\mathrm{rec}} \, \tilde{\phi}(E_{\mathrm{rec}}).
\end{equation}
The IACT likelihood is then computed assuming Gaussian uncertainties on the flux points:
\begin{equation}
    \mathcal{L}_{\mathrm{IACT}} = \prod_i \frac{1}{\sqrt{2\pi}\sigma_i} \exp\left(-\frac{(\phi_i^{\mathrm{obs}} - \phi_i^{\mathrm{pred}})^2}{2\sigma_i^2}\right),
\end{equation}
where $\phi_i^{\mathrm{obs}}$ and $\phi_i^{\mathrm{pred}}$ are the observed and predicted fluxes in analysis bin $i$, and $\sigma_i$ is the measurement uncertainty. Two approximations are implicit here and inherited from \citep{HAWC_ALP_2023}: the published flux points are treated as uncorrelated, and the energy dispersion is applied to points that are themselves unfolded under a spectral assumption.

\subsubsection{HAWC Likelihood}

For the HAWC data, we compute the Poisson likelihood using the Multi-Mission Maximum Likelihood framework (3ML)~\citep{threeML2015}, incorporating the nominal IRFs:
\begin{equation}
    \mathcal{L}_{\mathrm{HAWC}} = \prod_{i,j} \frac{\mu_{ij}^{n_{ij}^{\mathrm{obs}}} e^{-\mu_{ij}}}{n_{ij}^{\mathrm{obs}}!},
\end{equation}
where $i$ indexes the analysis bins, $j$ indexes the spatial pixels, $n_{ij}^{\mathrm{obs}}$ is the observed count, and $\mu_{ij} = S_{ij} + B_{ij}$ is the expected count (signal plus background).

\subsection{Statistical Method} \label{sec:stat}

\begin{figure*}[ht!]
    \centering
    \includegraphics[width=\linewidth]{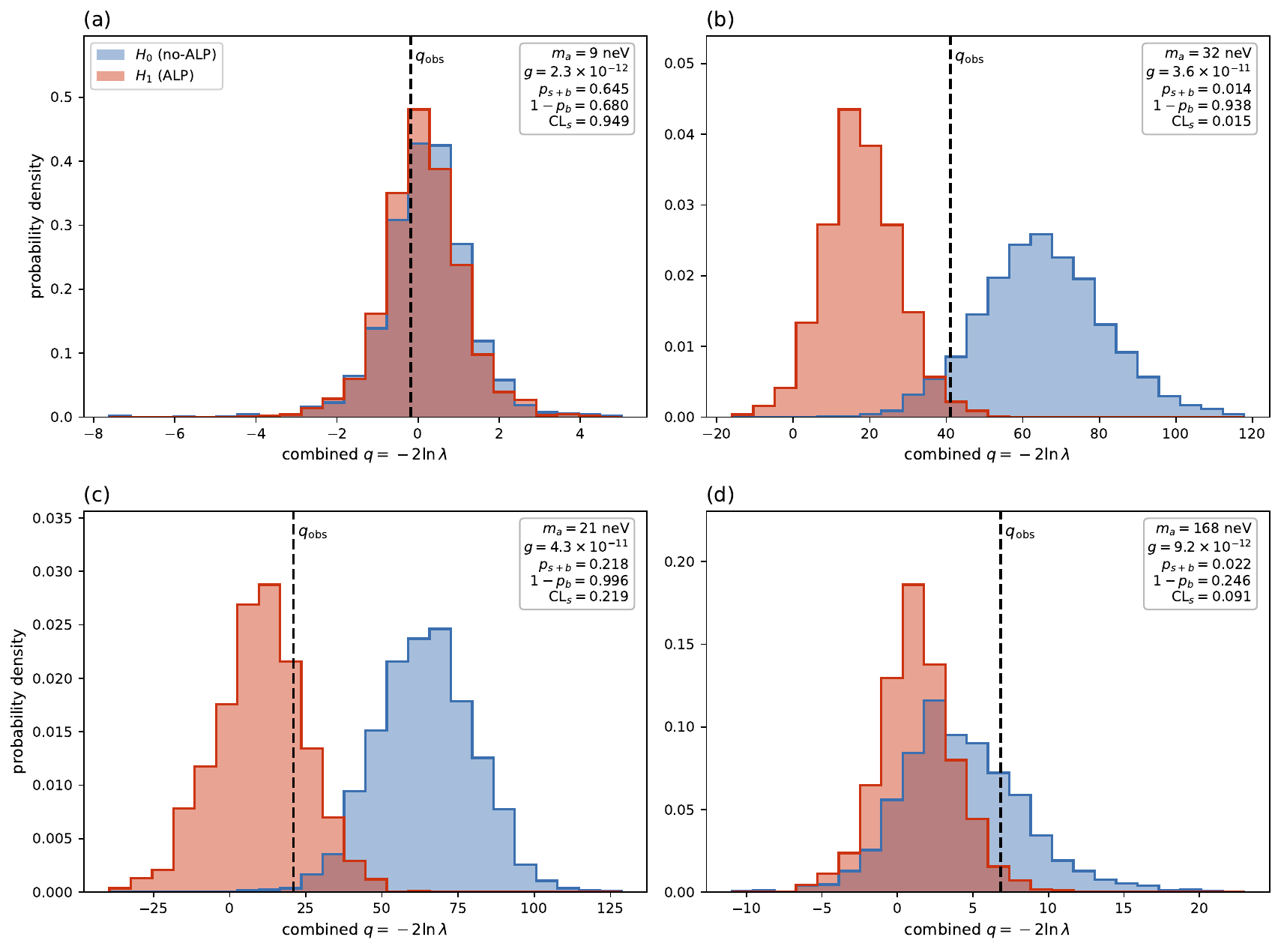}
    \caption{Distributions of the combined test statistic $q = -2\ln\lambda$ under the no-ALP ($H_0$, blue) and ALP ($H_1$, red) hypotheses, from the $2000$ pseudo-experiments, at four representative ALP candidates. The dashed line is the observed $q$. The inset lists $m_a$, $g_{a\gamma\gamma}$, the tail probabilities $p_{s+b} = P(q\ge q_{\mathrm{obs}}|H_1)$ and $1-p_b = P(q\ge q_{\mathrm{obs}}|H_0)$, and $\mathrm{CL}_s = p_{s+b}/(1-p_b)$. (a) A point outside the experimental reach: $H_0$ and $H_1$ overlap and $\mathrm{CL}_s \approx 1$. (b) An excluded point: the distributions separate and the observed $q$ is consistent with the no-ALP hypothesis. (c) A non-excluded ``hole'' embedded in the excluded region, where the observed $q$ remains compatible with the ALP hypothesis. (d) A point with $p_{s+b} < 0.05$ that is protected from exclusion by the CL$_s$ normalization, since the small $1-p_b$ reflects low sensitivity.}
    \label{fig:cls_hist}
\end{figure*}

We test the hypotheses on a coarser $25 \times 25$ grid covering $m_a \in [0.1, 2000]$~neV and $g_{a\gamma\gamma} \in [10^{-12}, 6 \times 10^{-11}]$~GeV$^{-1}$. The upper bound on the coupling is motivated by the CERN Axion Solar Telescope (CAST) limit~\citep{CAST2017}. The grid spacing is set by the cost of the CL$_s$ calibration: each point requires $2000$ Poisson-fluctuated ALP pseudo-experiments, with the intrinsic-spectrum parameters of all seven sources refit in each. At each point $P_{\gamma\gamma}$ is interpolated from the tabulation of Section~\ref{sec:pgg}, which extends beyond the scanned region so that boundary effects do not affect the scan.

The profile likelihood ratio test statistic, $q$, is computed as:
\begin{equation}
    q = -2 \ln \frac{\max_{\boldsymbol{\theta}} \mathcal{L}_{\mathrm{total}}(\mathrm{data} \,|\, m_a, g_{a\gamma\gamma}, \boldsymbol{\theta})}{\max_{\boldsymbol{\theta}} \mathcal{L}_{\mathrm{total}}(\mathrm{data} \,|\, m_a = g_{a\gamma\gamma} = 0, \boldsymbol{\theta})},
\end{equation}
where $\boldsymbol{\theta}$ represents nuisance parameters (the intrinsic flux normalization and spectral index). We write the likelihood ratio as $\lambda$, so that $q = -2\ln\lambda$. The asymptotic $\chi^2$ form of Wilks' theorem does not apply here. Under the no-ALP hypothesis the likelihood does not depend on $m_a$, so the mass is unidentified, and the coupling sits on the boundary of its allowed range at $g_{a\gamma\gamma} = 0$~\citep{Chernoff1954}. Either alone violates the regularity conditions of the theorem, and the strongly non-monotonic dependence of $P_{\gamma\gamma}$ on $g_{a\gamma\gamma}$ compounds them. 
The distribution of $q$ must therefore be obtained from Monte-Carlo pseudo-experiments rather than assumed. A fixed $q$-threshold criterion (e.g., $q > 5.99$, the $\chi^2$ value for two degrees of freedom at 95\%) is consequently not valid here \citep{FermiLAT2016, HAWC_M87_2026}.

We therefore employ the CL$_s$ method to establish conservative exclusion limits that explicitly account for the experimental sensitivity~\citep{Junk1999,Read2002}. We generate $N = 2000$ pseudo-experiments under each hypothesis. Because the no-ALP model does not depend on the ALP parameters, a single set of $2000$ null datasets is generated once and evaluated at every point of the scan, so that the maximum improvement (minimum $q$) over the grid is defined for each pseudo-experiment. 
The ALP ensembles are generated separately at each grid point from the ALP model at that point. Each pseudo-experiment consists of a Poisson fluctuation of the expected HAWC counts (source plus background) in every spatial pixel and analysis bin, together with a Gaussian fluctuation of the predicted IACT flux points within their measurement uncertainties. The generating models are the best-fit no-ALP model and the best-fit ALP model at the sampled grid point, each obtained by profiling $K$ and $\Gamma$ on the observed data, and $K$ and $\Gamma$ are refit for every pseudo-experiment. The same pseudo-experiment construction is described in \citep{2025icrc.confE.538W}.

The CL$_s$ value is computed as:
\begin{equation}
    \mathrm{CL}_s = \frac{p_{s+b}}{1 - p_b},
\end{equation}
where $p_{s+b} = P(q \geq q_{\mathrm{obs}}\,|\,H_1)$ is the probability, under the ALP hypothesis ($H_1$), of obtaining a test statistic at least as large as observed, and $p_b = P(q < q_{\mathrm{obs}}\,|\,H_0)$ is the probability under the no-ALP hypothesis ($H_0$), so that $1 - p_b = P(q \geq q_{\mathrm{obs}}\,|\,H_0)$. A parameter-space point is excluded at the 95\% $\mathrm{CL}_s$ level if $\mathrm{CL}_s < 0.05$. By normalizing $p_{s+b}$ by $1 - p_b$, the CL$_s$ method prevents a downward fluctuation of the data in a region of low sensitivity (where a bare $p_{s+b}$ criterion would exclude) from producing a spurious limit.

Figure~\ref{fig:cls_hist} illustrates four particular regimes of the combined test, using the $q$ distributions under $H_0$ and $H_1$ at four representative ALP candidates. In panel~(a) the experiment has no sensitivity: the two distributions overlap and $\mathrm{CL}_s \approx 1$. In panel~(b) the experiment is sensitive and the data are consistent with the no-ALP hypothesis, so the distributions separate and $\mathrm{CL}_s$ falls well below $0.05$. Panel~(c) shows an isolated non-excluded ``hole'' embedded in the excluded region, where the observed $q$ remains compatible with the ALP hypothesis. Section~\ref{sec:limits} shows that such holes are consistent with statistical fluctuations of the dataset. Panel~(d) shows a point with $p_{s+b} < 0.05$ that is nonetheless not excluded: the small $1-p_b$ reflects a downward fluctuation of the data in a region of low sensitivity, which the $\mathrm{CL}_s$ normalization corrects for.

\section{Results}

\subsection{ALP Exclusion Limits} \label{sec:limits}

\begin{figure}[ht!]
    \centering
    \includegraphics[width=\linewidth]{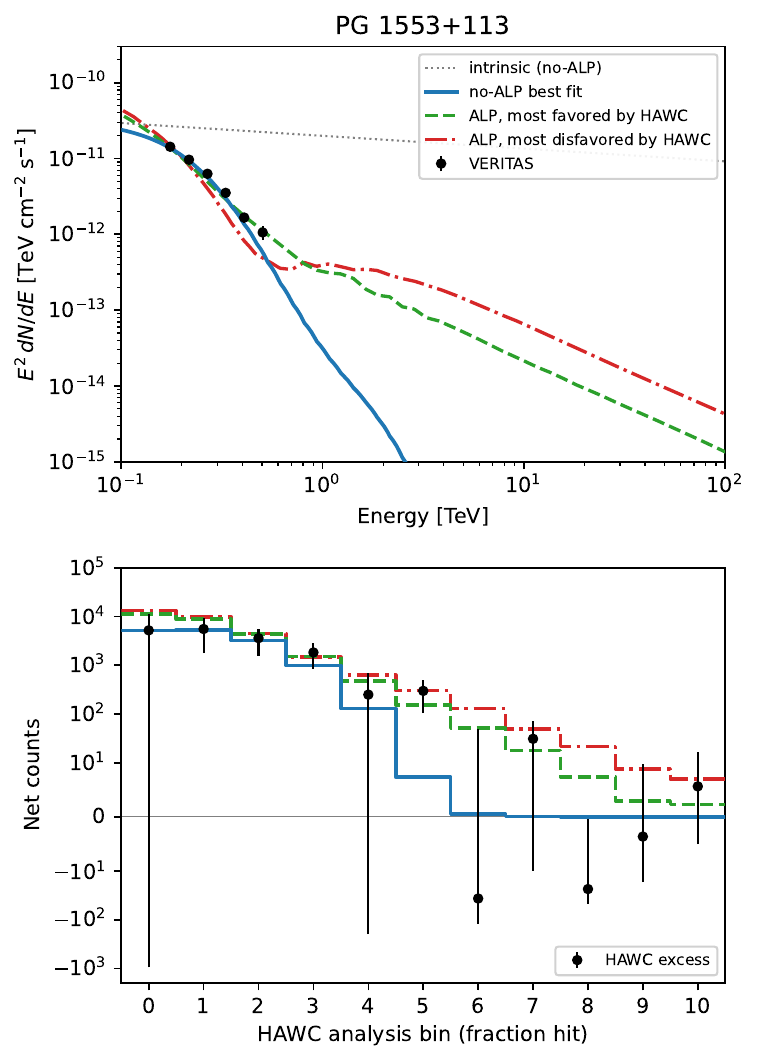}
    \caption{Observed spectrum and HAWC counts for PG~1553+113, the most constraining source in the sample (Section~\ref{sec:contrib}), compared with the no-ALP best fit (blue solid) and with the ALP hypothesis at the two grid points where the HAWC term of the observed test statistic of this source is smallest and largest. These are $m_a = 2.7$~neV, $g_{a\gamma\gamma} = 3.6 \times 10^{-11}$~GeV$^{-1}$, the point the HAWC data most favor (green dashed), and $m_a = 48.7$~neV, $g_{a\gamma\gamma} = 6.0 \times 10^{-11}$~GeV$^{-1}$, the point they most disfavor (red dot-dashed). Each ALP curve uses the intrinsic-spectrum parameters profiled at its own grid point. Top: spectral energy distribution with the published VERITAS flux points. The dotted line shows the unattenuated intrinsic spectrum of the no-ALP fit. Bottom: net excess counts within $0.5^\circ$ of the source position per HAWC fraction-hit analysis bin (on-array and off-array configurations summed), with $\sqrt{N_{\mathrm{obs}}}$ uncertainties. The histograms show the expected excess for the three models, with the same color coding as the top panel. In the lowest bins the point-spread function is substantially broader than the $0.5^\circ$ aperture, so the aperture contains only part of the source counts and the sums may include unmodeled field emission. The likelihood itself is evaluated per pixel.}
    \label{fig:spectra}
\end{figure}

\begin{figure*}[ht!]
    \centering
    \includegraphics[width=\linewidth]{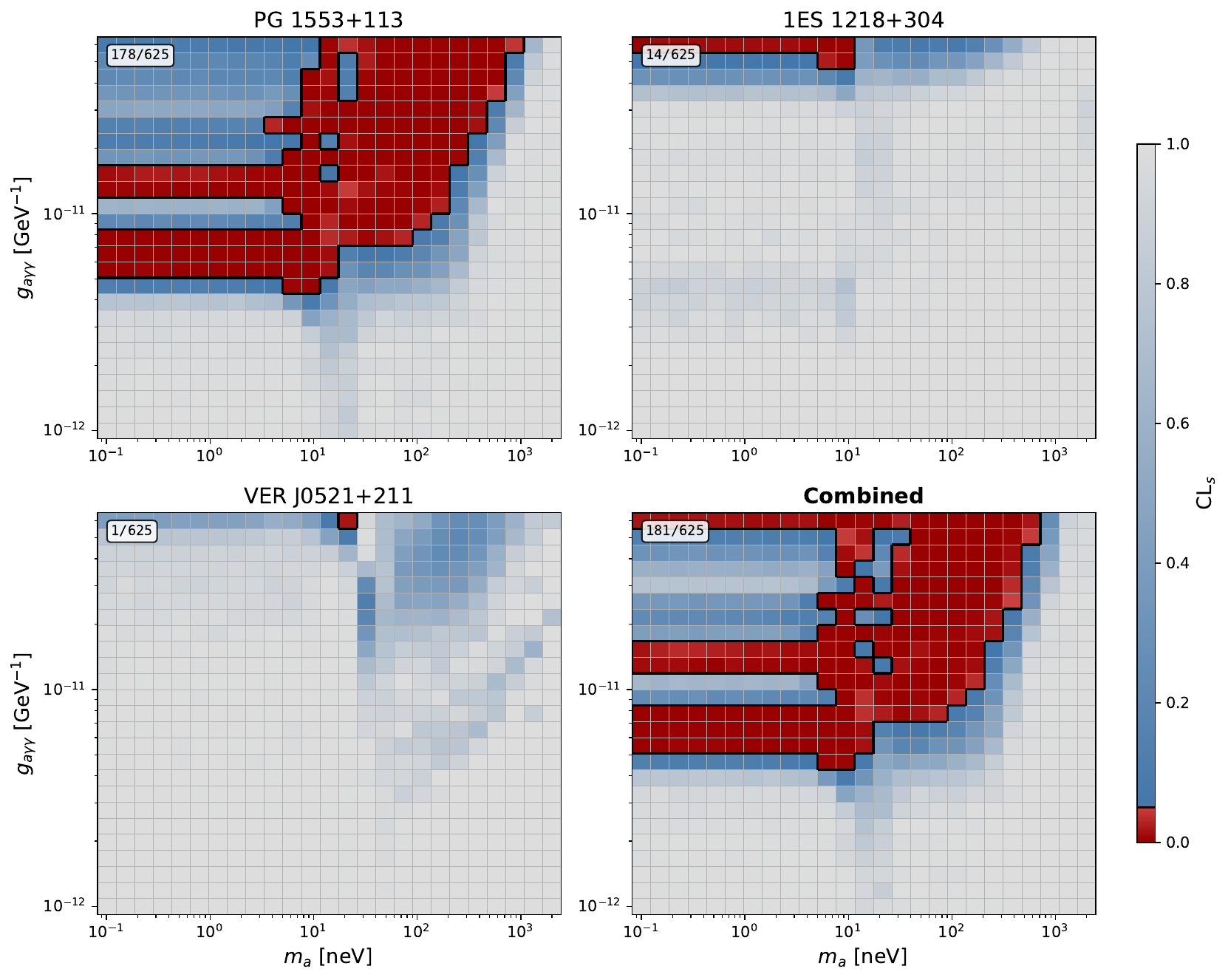}
    \caption{CL$_s$ over the scanned $(m_a, g_{a\gamma\gamma})$ grid for the three sources that individually set 95\% CL$_s$ limits (PG~1553+113, 1ES~1218+304, VER~J0521+211) and for the combined analysis. Each cell is colored by its CL$_s$ value. Red cells are excluded at 95\% CL$_s$ ($\mathrm{CL}_s < 0.05$), and the black line marks the grid-aligned exclusion boundary (excluded-cell count out of 625 inset in each panel). The combined analysis excludes 181/625 cells.}
    \label{fig:cls_surface}
\end{figure*}

\begin{figure*}[ht!]
    \centering
    \includegraphics[width=0.75\linewidth]{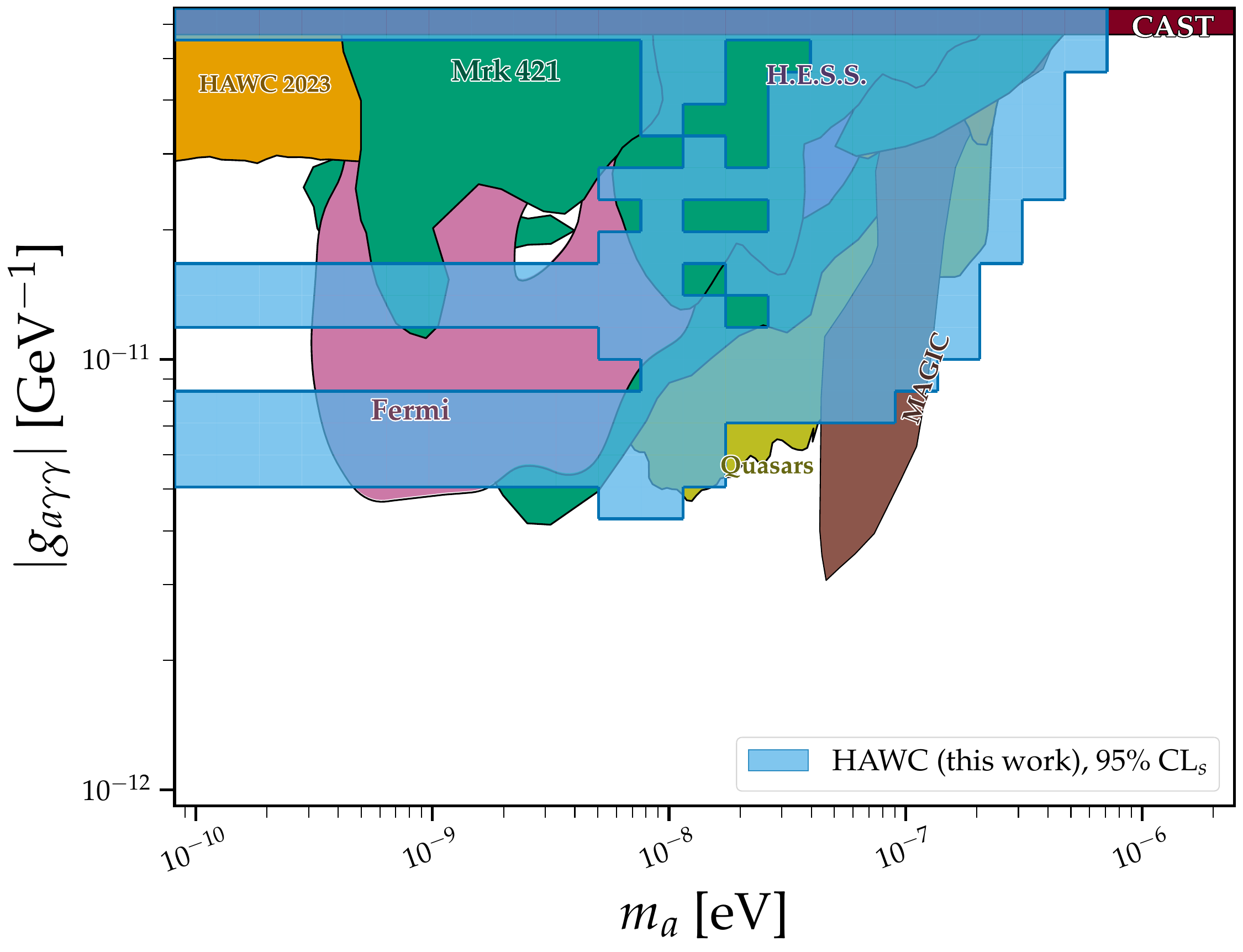}
    \caption{Combined 95\% CL$_s$ exclusion from the joint analysis of seven blazars, shown as the excluded cells of the scan grid as in Figure~\ref{fig:cls_surface}, compared with other limits on photon-ALP conversion in AGN, as compiled in \citep{AxionLimits}: the earlier HAWC blazar search~\citep{HAWC_ALP_2023}, Mrk~421~\citep{Mrk421ALP2021,Mrk421ALP2022,Mrk421ALP2024}, PKS~2155$-$304 with H.E.S.S.~\citep{HESS2013ALP}, the Perseus cluster with MAGIC~\citep{MAGIC2024ALP}, NGC~1275 with Fermi-LAT~\citep{FermiLAT2016}, and three flaring flat-spectrum radio quasars~\citep{FermiQuasars2023}. The CAST bound~\citep{CAST2017} sets the upper edge of the scanned coupling range. The non-excluded gaps and interior holes are part of the result (Section~\ref{sec:limits}).}
    \label{fig:alp_limits}
\end{figure*}

No statistically significant preference for an ALP signal is found. The ALP hypothesis is most favored at $m_a \approx 6$~neV and $g_{a\gamma\gamma} \approx 3 \times 10^{-12}$~GeV$^{-1}$, where it improves the fit over the no-ALP model by $2\ln(\mathcal{L}_{\mathrm{ALP}}/\mathcal{L}_0) = 1.7$. Since Wilks' theorem does not apply (Section~\ref{sec:stat}), we compare this value with the distribution of the maximum improvement over the grid in the no-ALP pseudo-experiments. The maximization over all $625$ grid points accounts for the look-elsewhere effect~\citep{GrossVitells2010}. The global significance is $0.5\sigma$ ($p = 0.31$). We therefore set exclusion limits.

Figure~\ref{fig:spectra} shows the HAWC contribution for PG~1553+113, the source dominating the combined constraint (Section~\ref{sec:contrib}). The IACT and HAWC likelihood terms are recorded separately in the scan, so that the HAWC contribution to the test statistic is available at every grid point. The two ALP curves correspond to the grid points at which this contribution is smallest and largest for this source.

Figure~\ref{fig:cls_surface} maps the resulting CL$_s$ values across the $(m_a, g_{a\gamma\gamma})$ parameter space for individual sources and the combined analysis.
The combined CL$_s$ is computed from the distributions of $q_{\mathrm{combined}}$ (Section~\ref{sec:analysis}) generated under the no-ALP and ALP hypotheses across all sources simultaneously.
The combined 95\% CL$_s$ excluded region is shown in Figure~\ref{fig:alp_limits}, placed in the context of existing astrophysical and laboratory constraints.
Because the photon survival probability oscillates as a function of the coupling as well as of the energy, the excluded set need not be simply connected, and here it is not. We therefore quote the limit at each mass as the lowest excluded coupling, and not as the statement that all larger couplings are excluded. This differs from the usual convention, in which every coupling above the quoted curve is excluded. The primary result is the excluded set of Figures~\ref{fig:cls_surface} and~\ref{fig:alp_limits}; the curve quoted below is its lower edge and coincides with a conventional upper bound only where the excluded region is contiguous, i.e.\ for $m_a \gtrsim 50$~neV. The fraction of excluded couplings above this edge is given separately. Defined this way, the joint analysis excludes couplings down to $g_{a\gamma\gamma} = 5.5 \times 10^{-12}$~GeV$^{-1}$ for $m_a \lesssim 4$~neV, reaching a minimum of $4.6 \times 10^{-12}$~GeV$^{-1}$ near $m_a \approx 6$--$9$~neV where the analysis is most sensitive, with the excluded region extending to $m_a \sim 600$~neV at progressively larger couplings. Both values coincide with the median expectation obtained from the null pseudo-experiments on the same grid.

The excluded fraction of the region above this boundary depends on the ALP mass. For $m_a \gtrsim 50$~neV every scanned coupling above the boundary is excluded, so the limit there has the conventional meaning of an upper bound on $g_{a\gamma\gamma}$. At $m_a \lesssim 4$~neV, by contrast, the exclusion breaks into bands separated by non-excluded coupling intervals, and only $40\%$ of the scanned cells above the boundary are excluded.

These limits also bear on the ALP interpretation of the reported opacity anomalies (Section~\ref{sec:intro}). Explaining the anomalies through photon-ALP conversion requires couplings ranging from $\sim 10^{-12}$~GeV$^{-1}$ in optimistic magnetic-field scenarios to $\sim 2 \times 10^{-11}$~GeV$^{-1}$ under conservative assumptions~\citep{Meyer2013}. Couplings at the conservative end of this range are excluded by the present analysis for $m_a \approx 30$--$300$~neV, while the optimistic end lies below the current sensitivity.

\subsection{Individual Source Contributions} \label{sec:contrib}

As shown in Figure~\ref{fig:cls_surface}, three of seven blazars (PG~1553+113, 1ES~1218+304, and VER~J0521+211) independently establish 95\% CL$_s$ exclusion limits. These are precisely the three sources that HAWC detects significantly on its own. Evaluating the HAWC likelihood alone for the no-ALP model, with the spectral index fixed to the joint best fit and the normalization profiled, gives $\sqrt{\mathrm{TS}}_{\mathrm{HAWC}} = 6.6$, $5.9$, and $5.7$ for 1ES~1218+304, VER~J0521+211, and PG~1553+113. The remaining four sources reach at most $\sqrt{\mathrm{TS}}_{\mathrm{HAWC}} = 2.6$ (Table~\ref{tab:sources_coords}). A significant HAWC detection is what allows the multi-TeV flux to be measured rather than merely bounded, and it is therefore the primary requirement for a source to constrain the ALP hypothesis on its own.

None of these sources was detected by HAWC in the analysis of \citep{HAWC_ALP_2023}, which reported signed significances between $-1.6$ and $2.24$ from the 3HWC catalog, negative values corresponding to a best-fit flux below zero. 1ES~1218+304 was the most significant of that sample as well. The increase to $\sqrt{\mathrm{TS}}_{\mathrm{HAWC}} = 6.6$ for that source follows from two effects of comparable size. 
The 3HWC catalog~\citep{3HWC} covers 2014 November to 2019 June ($1{,}523$~days of livetime), whereas the present dataset extends to 2024 January ($\sim$$3{,}000$~days); the doubled exposure corresponds to a factor $\sqrt{2}$ in significance.
We attribute the remaining factor of about two to the multilayer-perceptron gamma-hadron separation combined with the forward-folded fit to all 22 analysis bins, in place of a single flux estimate at 7~TeV evaluated for a fixed spectral index.

Among the three, PG~1553+113 provides the dominant contribution to the combined constraint. Its HAWC detection is comparable to those of the other two, but its much higher redshift ($z = 0.491$, against $0.182$ and $0.18$) implies far more severe EBL attenuation at TeV energies, so the ALP-induced transparency signature it would produce is correspondingly larger. Its magnetic field environment (jet + ICMF + GMF) also provides more conversion regions.

The remaining four sources (1ES~0229+200, 1ES~1011+496, 3C~66A, and PKS~1424+240) do not individually yield 95\% CL$_s$ exclusions, and their contribution to the joint result is small. The combined seven-source exclusion covers 181 of 625 grid cells against 178 for PG~1553+113 alone, a difference well inside the $26\%$ statistical scatter of that quantity (Section~\ref{sec:syst}). The combined constraint is therefore set almost entirely by PG~1553+113.

The dominance of PG~1553+113 does not make the combination one-sided. The sample was fixed in advance to that of \citep{HAWC_ALP_2023}, and every source enters the test statistic with full weight, pulling in both directions: the net contribution of the other six sources is negative on 41\% of the excluded cells, and ten cells that PG~1553+113 alone would exclude are not excluded in the combination, while thirteen others are added. The quoted limit is therefore the aggregate of a pre-defined sample, not the selection of its most constraining member.

\subsection{Comparison with Other Experiments}

Figure~\ref{fig:alp_limits} compares the present result with other limits on photon-ALP conversion in AGN. The comparison is restricted to this source class, whose limits rest on the same conversion environments (blazar jet, intra-cluster field, GMF) and the same modeling uncertainties. The upper boundary of the scanned coupling range is set by the CAST limit, $g_{a\gamma\gamma} < 6.6 \times 10^{-11}$~GeV$^{-1}$ for $m_a \lesssim 0.02$~eV~\citep{CAST2017}. Fermi-LAT searches for spectral irregularities in the gamma-ray spectra of active galactic nuclei (AGN) probe a similar mass range but are subject to distinct systematic uncertainties inherent to the GeV energy range~\citep{FermiLAT2016}.  
The present analysis reaches comparable couplings over a broader mass range with independent systematics. Previous IACT analyses have established limits from the spectra of individual sources: H.E.S.S. from the irregularity of the PKS~2155$-$304 spectrum~\citep{HESS2013ALP}, and MAGIC from NGC~1275 in the Perseus cluster~\citep{MAGIC2024ALP}. Limits from Mrk~421 have been derived by combining Fermi-LAT data with published flux points from ARGO-YBJ, MAGIC, and HAWC~\citep{Mrk421ALP2021,Mrk421ALP2022,Mrk421ALP2024}, and from a joint analysis of three flaring flat-spectrum radio quasars~\citep{FermiQuasars2023}. The present work extends these constraints to higher energies using HAWC.  
The most direct comparison is with the earlier HAWC blazar search of \citep{HAWC_ALP_2023}, which analyzed the same seven sources and excluded $g_{a\gamma\gamma} \gtrsim 3 \times 10^{-11}$~GeV$^{-1}$ for $m_a \lesssim 300$~neV. The two limits differ in confidence level, in the calibration of the test statistic, and in the definition of the excluded set (Section~\ref{sec:analysis}). 

X-ray limits from galaxy clusters~\citep{Reynolds2020} and supernova bounds are omitted, since they probe conversion in cluster cores and ALP production in stellar cores, respectively, rather than propagation along blazar lines of sight.

\subsection{Systematic Uncertainties} \label{sec:syst}

\begin{figure*}[ht!]
    \centering
    \includegraphics[width=\linewidth]{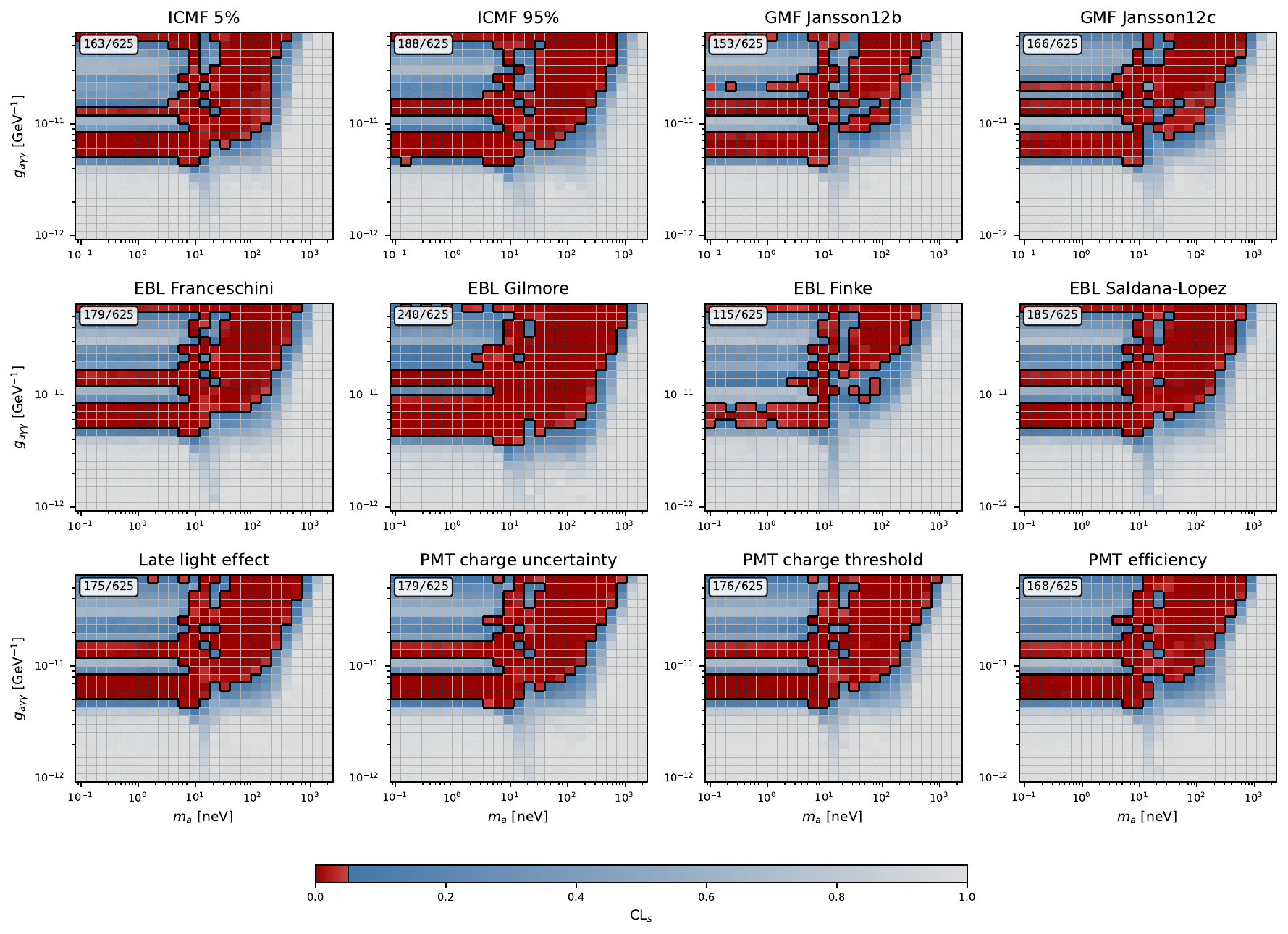}
    \caption{Systematic variations of the seven-source combined CL$_s$ over the scanned $(m_a, g_{a\gamma\gamma})$ grid. Top row: ICMF turbulence quantiles (5\%, 95\%) and GMF model variations (Jansson12b, Jansson12c). Middle row: alternative EBL models (Franceschini, Gilmore, Finke, Saldana-Lopez). Bottom row: the largest-deviation detector-response variant in each category (late light effect, PMT charge uncertainty, PMT charge threshold, PMT efficiency). In each panel, red cells are excluded at 95\% CL$_s$ ($\mathrm{CL}_s < 0.05$), the black line traces the grid-aligned boundary of the excluded region, and the inset reports the excluded-cell count out of 625 (nominal: 181/625).}
    \label{fig:systematics}
\end{figure*}

We evaluate systematic uncertainties arising from four primary sources: HAWC detector response, GMF modeling, EBL modeling, and ICMF modeling. Following the methodology of \citep{FermiLAT2016}, the impact of each systematic effect is quantified by computing the fractional change in the excluded area of the ALP parameter space:
\begin{equation}
    \varepsilon = \frac{\mathrm{Area}_{\mathrm{sys}} - \mathrm{Area}_{\mathrm{nom}}}{\mathrm{Area}_{\mathrm{nom}}} \times 100\%.
\end{equation}
These evaluated systematic effects are summarized in Table~\ref{tab:systematic_uncertainty} and visually represented in Figure~\ref{fig:systematics}.

\begin{table}[ht!]
\caption{Systematic uncertainties on the excluded ALP parameter-space area.}
\label{tab:systematic_uncertainty}
\begin{ruledtabular}
\begin{tabular}{lc}
Systematic Source & $\varepsilon$ (\%) \\
\hline
\multicolumn{2}{c}{\textit{Detector Response}} \\
\hline
Late light effect & $-3.3$ \\
PMT charge uncertainty & $-1.1$ \\
PMT charge threshold & $-2.8$ \\
PMT efficiency & $-7.2$ \\
\hline
\multicolumn{2}{c}{\textit{Environment Modeling}} \\
\hline
GMF modeling & $-15.5$ \\
EBL modeling & $-36.5$, $+32.6$ \\
ICMF quantiles & $-9.9$, $+3.9$ \\
\end{tabular}
\end{ruledtabular}
\end{table}

\subsubsection{Detector Response}

Four systematic effects related to the HAWC detector response are evaluated by systematically varying the IRFs: late light effect (delayed optical photons affecting timing/charge reconstruction), PMT charge calibration uncertainty, PMT hit detection thresholds, and PMT quantum efficiency variations. For each category, multiple IRF variations are tested and the most significant deviation from nominal is reported. The detector systematics all remain below the 10\% level, with the per-run PMT efficiency ($-7.2\%$) and the late light effect ($-3.3\%$) giving the largest deviations.

\subsubsection{Environment Modeling}

GMF modeling uncertainties are assessed by comparing the baseline Jansson12a configuration to parametric variations~\citep{JanssonFarrar2012}. The Jansson12 model describes the GMF as a superposition of a disk field, a toroidal halo field, and an X-shaped poloidal halo field, with parameters constrained by Faraday rotation measures and synchrotron polarization data. The alternative configurations tested are Jansson12b and Jansson12c, the parameter sets re-fitted to Planck dust polarization data~\citep{Planck2016GMF}. These change the excluded area by $-15.5\%$ and $-8.3\%$, respectively. We restrict the GMF systematic to parametric variations of the Jansson12 family, and the reported estimate is representative of this model ensemble.

The EBL modeling uncertainty is assessed using alternative prescriptions to the baseline \citep{Dominguez2011} model. Four are considered: the empirical model of \citep{Franceschini2008}, based on galaxy luminosity functions and spectral energy distributions; the semi-analytical model of \citep{Gilmore2012}, using $\Lambda$CDM cosmology and stellar population synthesis; the model of \citep{Finke2022}, based on a self-consistent calculation of the cosmic star-formation history and dust emission; and the empirical model of \citep{SaldanaLopez2021}, constrained by Fermi-LAT observations of blazars. These models predict different EBL photon densities as a function of wavelength and redshift, leading to variations in the gamma-ray optical depth $\tau(E, z)$. The EBL systematic is the largest single contribution, ranging from $-36.5\%$ (Finke) to $+32.6\%$ (Gilmore). Since the analysis fits the EBL-convolved spectrum with a free flux normalization and spectral index, this sensitivity reflects the differing energy dependence (shape) of the optical-depth curve across models rather than the overall opacity level.

The ICMF modeling introduces variations in the photon survival probability through the stochastic Gaussian turbulence. For sources with ICMF components, this is evaluated using the 5\%, 50\% (baseline), and 95\% quantiles of the distribution from $N_{\mathrm{sim}} = 1000$ turbulence realizations, and contributes $-9.9\%$ to $+3.9\%$.

Among the evaluated effects, the EBL modeling dominates the systematic budget ($-36.5\%$ to $+32.6\%$), with subdominant contributions from the GMF ($-15.5\%$) and ICMF ($-9.9\%$ to $+3.9\%$) modeling. The HAWC detector response remains the smallest category, with the largest single variation of $-7.2\%$ from the PMT efficiency.

\section{Summary and Conclusions}

A search for axion-like particles has been performed using a joint likelihood analysis of HAWC and IACT observations of seven BL Lac blazars spanning redshifts $z = 0.14$--$0.60$. Approximately 3,000~days of HAWC Pass 5 observations, combined with published spectral measurements from H.E.S.S., MAGIC, and VERITAS, are used to probe the ALP parameter space over $m_a = 0.1$--$2000$~neV. 

Couplings are excluded at the 95\% CL$_s$ level down to $g_{a\gamma\gamma} = 5.5 \times 10^{-12}$~GeV$^{-1}$ for $m_a \lesssim 4$~neV and down to $4.6 \times 10^{-12}$~GeV$^{-1}$ near $m_a \approx 6$--$9$~neV, where the analysis is most sensitive. Varying the environment and detector-response models changes the excluded parameter-space area by $-36.5\%$ to $+32.6\%$. The excluded region is not simply connected: it is contiguous above the exclusion boundary for $m_a \gtrsim 50$~neV but breaks into bands at the lowest masses, where the pattern of gaps is consistent with statistical fluctuation of the data. Three sources individually exclude regions of ALP parameter space at 95\% CL$_s$: PG~1553+113, 1ES~1218+304, and VER~J0521+211. The systematic uncertainty on the excluded parameter space area is dominated by the EBL modeling ($-36.5\%$ to $+32.6\%$), with subdominant contributions from the GMF and ICMF modeling and detector-response effects below the 10\% level. The excluded region also covers, for $m_a \approx 30$--$300$~neV, the couplings required under conservative magnetic-field assumptions to explain the reported VHE opacity anomalies~\citep{Meyer2013}.

This analysis extends the methodology of \citep{HAWC_ALP_2023} by employing the full HAWC IRFs and a Poisson likelihood, rather than Gaussian approximations based on public flux estimates, and improves the earlier limit by a factor of about five at the lowest ALP masses and about six near $m_a \approx 6$--$9$~neV. 

Future observations can extend this search. The Cherenkov Telescope Array Observatory will improve the spectral measurements that anchor the intrinsic source spectra~\citep{CTA2019}, and the Southern Wide-field Gamma-ray Observatory would enlarge the accessible blazar sample~\citep{SWGOScience}. Improved determinations of the EBL and of the Galactic and intra-cluster magnetic fields would reduce the modeling uncertainties that dominate the present systematic budget. 

\begin{acknowledgments}
We acknowledge the support from: the US National Science Foundation (NSF); the US Department of Energy Office of High-Energy Physics; the Laboratory Directed Research and Development (LDRD) program of Los Alamos National Laboratory; Secretar\'{\i}a de Ciencia Humanidades Tecnolog\'{\i}a e Innovaci\'{o}n (SECIHTI), M\'{e}xico, grants LNC-2023-117, 271051, 232656, 260378, 179588, 254964, 258865, 243290, 132197, A1-S-46288, A1-S-22784, CF-2023-I-645, CBF-2023-2024-1630, CBF-2025-G-73, C\'{a}tedras 873, 1563, 341, 323, Red HAWC, M\'{e}xico; DGAPA-UNAM grants IG101323, IG100726, IN111716-3, IN111419, IA102019, IN106521, IN114924, IN110521, IN102223; VIEP-BUAP; PIFI 2012, 2013, PROFOCIE 2014, 2015; the University of Wisconsin Alumni Research Foundation; the Institute of Geophysics, Planetary Physics, and Signatures at Los Alamos National Laboratory; Polish Science Centre grant, 2024/53/B/ST9/02671; Coordinaci\'{o}n de la Investigaci\'{o}n Cient\'{\i}fica de la Universidad Michoacana; Royal Society - Newton Advanced Fellowship 180385; Gobierno de Espa\~{n}a and European Union-NextGenerationEU, grant CNS2023-144099; The Program Management Unit for Human Resources \& Institutional Development, Research and Innovation, NXPO (grant number B16F630069); Coordinaci\'{o}n General Acad\'{e}mica e Innovaci\'{o}n (CGAI-UdeG), PRODEP-SEP UDG-CA-499; Institute of Cosmic Ray Research (ICRR), University of Tokyo. H.M. acknowledges support under grant number CBF2023-2024-1630. H.F. acknowledges support by NASA under award number 80GSFC21M0002. C.R. acknowledges support from National Research Foundation of Korea (RS-2023-00280210, RS-2026-25470489). J.L. acknowledges support from National Research Foundation of Korea (RS-2023-NR076954). Y.S. acknowledges support from National Research Foundation of Korea (RS-2026-25556928). I.W. acknowledges support from National Research Foundation of Korea (RS-2021-NR058944). We also acknowledge the significant contributions over many years of Stefan Westerhoff, Gaurang Yodh and Arnulfo Zepeda Dom\'{\i}nguez, all deceased members of the HAWC collaboration. Thanks to Scott Delay, Luciano D\'{\i}az and Eduardo Murrieta for technical support.
\end{acknowledgments}

\bibliography{main}

\end{document}